\documentclass[fleqn,usenatbib]{mnras} 

\usepackage{newtxtext,newtxmath}
\usepackage[normalem]{ulem}
\usepackage{soul}
\usepackage{xcolor}

\usepackage[T1]{fontenc}

\DeclareRobustCommand{\VAN}[3]{#2}
\let\VANthebibliography\thebibliography
\def\thebibliography{\DeclareRobustCommand{\VAN}[3]{##3}\VANthebibliography}

\newcommand{\ltsima}{$\; \buildrel < \over \sim \;$}
\newcommand{\lsim}{\lower.5ex\hbox{\ltsima}}
\newcommand{\gtsima}{$\; \buildrel > \over \sim \;$}
\newcommand{\gsim}{\lower.5ex\hbox{\gtsima}}

\usepackage{graphicx}	
\graphicspath{{./}{figures/}}
\usepackage{amsmath}	

\newcommand{\hypa}{\texttt{IPMSim}}
\newcommand{\Msun}{M$_{\odot}$}

\newcommand{\Ngal}{$N=32$}
\newcommand{\msun}{{\rm M}_\odot}
\newcommand{\Rv}{$R_{\rm v}$}
\newcommand{\Mv}{M_{\rm v}}

\title[Quenching in Cosmic Sheets]{Quenching in Cosmic Sheets: Tracing the Impact of Large Scale Structure Collapse on the Evolution of Dwarf Galaxies}

\author[Imad Pasha \& Nir Mandelker et al.]{Imad Pasha,$^{1}$\thanks{E-mail: imad.pasha@yale.edu}
Nir Mandelker,$^{2,3,1,4}$
Frank C. van den Bosch,$^{1}$
Volker Springel,$^{5}$\newauthor
 and Freeke van de Voort$^{6}$
\\
$^{1}$Department of Astronomy, Yale University, PO Box 208101, New Haven, CT, USA\\
$^{2}$Racah Institute of Physics, The Hebrew University of Jerusalem, Jerusalem 91904, Israel\\
$^{3}$Kavli Institute for Theoretical Physics, University of California, Santa Barbara, CA 93106, USA\\
$^{4}$Heidelberger Institut f{\"u}r Theoretische Studien, Schloss-Wolfsbrunnenweg 35, 69118 Heidelberg, Germany\\
$^{5}$Max Planck Institute for Astrophysics, Karl-Schwarzschild-Strasse 1, D-85748 Garching, Germany\\
$^{6}$Cardiff Hub for for Astrophysics Research and Technology, School of Physics and Astronomy, Cardiff University, Queen’s Buildings, The Parade, Cardiff CF24 3AA, UK
}

\date{Accepted XXX. Received YYY; in original form ZZZ}

\pubyear{2022}

\begin{document}
\label{firstpage}
\pagerange{\pageref{firstpage}--\pageref{lastpage}}
\maketitle

\begin{abstract}
Dwarf galaxies are thought to quench primarily due to environmental processes most typically occurring in galaxy groups and clusters or around single, massive galaxies. However, at earlier epochs, ($5 < z < 2$), the collapse of large scale structure (forming Zel'dovich sheets and subsequently filaments of the cosmic web) can produce volume-filling accretion shocks which elevate large swaths of the intergalactic medium (IGM) in these structures to a hot ($T>10^6$ K) phase. We study the impact of such an event on the evolution of central dwarf galaxies  ($5.5 < \log M_* < 8.5$) in the field using a spatially large, high resolution cosmological zoom simulation which covers the cosmic web environment between two protoclusters. We find that the shock-heated sheet acts as an environmental quencher much like clusters and filaments at lower redshift, creating a population of quenched, central dwarf galaxies. Even massive dwarfs which do not quench are affected by the shock, with reductions to their sSFR and gas accretion. This process can potentially explain the presence of isolated quenched dwarf galaxies, and represents an avenue of pre-processing, via which quenched satellites of bound systems quench before infall. 
\end{abstract}




\section{Introduction}

Quenching is the process by which galaxies, driven by either internal or external mechanisms (or both), cease forming stars. Understanding quenching mechanisms and how they affect galaxies in different mass regimes and environments is key to forming a coherent picture of galaxy evolution across cosmic time. It is generally accepted that massive galaxies in halos with $\Mv\gsim 10^{12}$ M$_{\odot}$ have the ability to self-quench, due to their creation and maintenance of a hot gaseous halo via feedback from supernovae and active galactic nuclei (AGN), combined with gravitational heating and the formation of a virial accretion shock, which together prevent the necessary accretion of cold gas from the intergalactic medium (IGM) \citep[e.g.,][]{Birnboim:2003,Keras:2005,Croton:2006,Dekel:2006,Dekel:2008}. While this model predicts the quenching of galaxies with $M_{*}\gtrsim$ $10^{10}$ M$_{\odot}$, galaxies below roughly this mass threshold are not expected to be able to produce and maintain a hot halo, and hence lack a mechanism for self quenching \citep{Gabor:2012}. While supernova feedback can eject gas from these galaxies and temporarily shut-off star-formation, the lack of a hot halo means that cold gas can will eventually still accrete onto the galaxy, from both the IGM and recycled ejecta, reigniting star-formation in what is sometimes referred to as a ``breathing mode'' \citep[e.g.][]{Dekel:1986,Stinson:2007}. As a result, galaxies in the mass range $10^6<M_{*}<10^9$ M$_{\odot}$ are thought to quench primarily environmentally, by entering a region of the universe that is non-conducive to the accretion of pristine cold gas and which may remove bound gas from embedded systems \citep[e.g.,][]{vdBosch:2008,Peng:2010,Peng:2012,Woo:2013,Wang:2018}. The most-commonly discussed environments of this type include the hot haloes of massive galaxies, as well as larger scale hot-phase intra-group and intra-cluster gas present in galaxy groups and clusters. In such environments, i.e., when a dwarf galaxy is a \textit{satellite}, processes such as ram pressure stripping \citep{Gunn:1972}, harassment, and strangulation \citep{Larson:1980} can strip the dwarf of its circumgalactic medium (CGM), prevent new gas from accreting, and ultimtely lead to the shutting off of star formation. 

In contrast, dwarf galaxies that are \textit{centrals} --- defined as being the primary galaxy in the dark matter halo with which they are associated in a group catalog \citep[e.g.,][]{Yang:2007} --- are expected to be star forming, as their designation as non-satellites implies they are field systems outside the hot environments discussed above. This prediction agrees well with studies that have shown a negligible quenched fraction among isolated dwarf galaxies \citep[e.g.,][]{Wang.etal.09, Geha:2012}.

Less often discussed in the context of environmental quenching are cosmological-scale accretion shocks created by the collapse of large scale structure. These shocks potentially mimic the environmental conditions commonly seen in hot haloes of massive galaxies, groups, and clusters, but over vastly larger spatial scales. Such accretion shocks are predicted theoretically \citep[e.g.][]{Mo:2010,Birnboim:2016}, and have also been identified in cosmological simulations around cosmic ``sheets" (also known as Zel'dovich sheets or pancakes) and filaments, particularly when sheets or filaments collide with each other \citep{Mandelker:2019,Ramsoy:2021}. Dwarf galaxies which happen to live in such environments  during (or enter after) such accretion shock-generating events experience a change in environment, in which their virial temperature,
\begin{equation}
\label{eqn:Tvir}
    T_{\rm V} \simeq 5.6 \times 10^4 \hspace{3pt}{\rm K}\hspace{3pt} \left(\frac{\mu}{0.59}\right)\left(\frac{ M_{\rm halo}}{10^{10}\hspace{2pt}\rm M_{\odot}}\right)^{2/3}\left(\frac{1+\mathit{z}}{4}\right),
\end{equation}
rapidly becomes comparable to, or significantly lower than, the ambient temperature of their surrounding medium. In this case, one would predict a reduction in their ability to accrete and retain gas, potentially leading to a shutoff of star-formation. Ram pressure as the galaxies plow through the hot medium can serve to further strip them of gas and reduce their star-forming ability. Thus, it is reasonable to expect that isolated, central dwarf galaxies, which are far from any massive halo but are present in cosmic sheets or filaments when such accretion shocks form, should experience similar quenching processes as satellites would from their host systems --- though it is important to note that large sheet regions are themselves heterogeneous environments. 

Indeed, some observational evidence at low redshift suggests the ``pre-processing" of galaxies that fall into clusters along filaments --- these systems appear to be more quenched than those that enter from off-filament \citep[e.g.,][]{Salerno:2020,Santiago-Bautista:2020}. Recent work analyzing the position of galaxies in SDSS relative to cosmic web features (e.g., filaments, walls, nodes) also finds evidence of gradients in galaxy properties, with suppressed SFRs nearer to cosmic structures \citep{Winkel:2021}. This quenching mechanism has also been suggested by simulations at low redshift \citep[e.g.,][]{Benitez:2013}, focusing in particular on ``fly-through" galaxies which quench via ram pressure interactions with cosmic pancakes. However, the generality of this class of phenomenon, as well as its redshift dependence, remains unknown. Further motivating such studies, the general suite of effects that components of the cosmic web (e.g., sheets and filaments) have on galaxies is not yet well understood, and there is conflicting evidence regarding whether such structures enhance star formation \citep[e.g.,][]{Vulcani:2019} or suppress it \citep[e.g.,][]{Kraljic:2018}, and indeed, some evidence suggests such effects might evolve or even trade places with cosmic time \citep[e.g.][]{Dekel:2006, Birnboim:2016}.

Thus, the evolution of central dwarf galaxies in large scale accretion shocks is worth further investigation, in particular during ``cosmic noon'', at $z\sim$ 2-4, when structure formation is at its peak. Due to the small sizes and low surface brightnesses of dwarf galaxies, however, it is intractable to observationally study their evolution within such cosmic environments at high redshift. Furthermore, observations are limited in their ability to constrain quenching mechanisms as quenching histories of systems must be inferred from single-epoch observations.

\begin{figure*}
    \centering
    \includegraphics[width=0.96\textwidth]{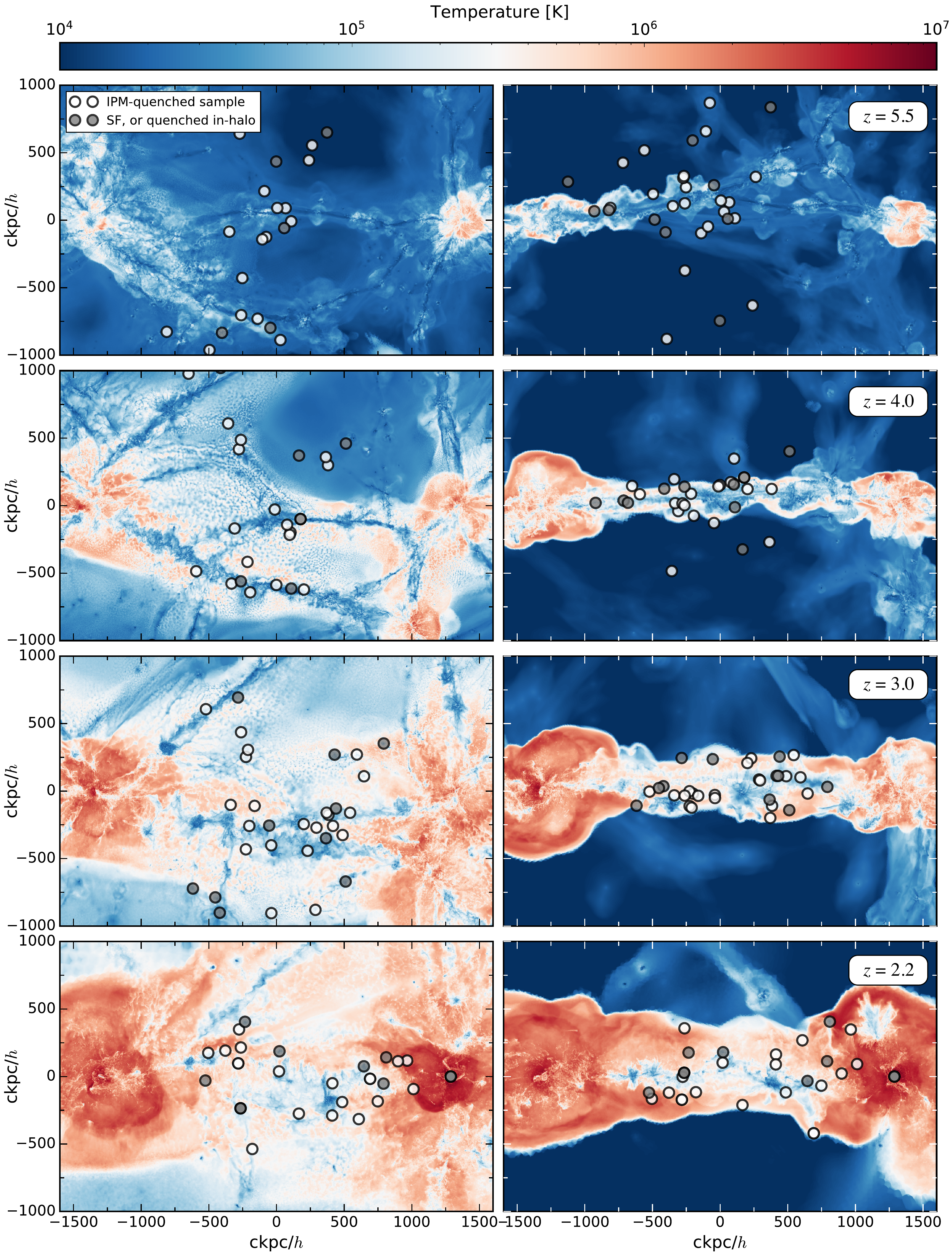}
    \caption{Face-on (left) and edge-on (right) views of the density-weighted temperature in the \hypa{} simulation at 4 different redshifts, with an integration depth of 400 ckpc/$h$. Marked in all panels are the galaxies selected at $z=3$ for analysis (see Section 2.2); cospatiality with the sheet at this redshift was a selection criterion, visible in the right hand panel. Points are colored by whether the given system quenches due to the accretion shock in the IPM (white) and those which either remain star forming, or quench within 3 \Rv{} of the proto-clusters (grey). }
    \label{fig:fullmaps}
\end{figure*}

Cosmological hydrodynamical simulations thus provide the best current method for tracking the evolution of low-mass galaxies and their interaction with collapsing large-scale structure and the cosmic web. However, studying this process in detail is a challenge even with current cosmological simulations. Large volume cosmological simulations such as EAGLE \citep{Schaye:2015}, Illustris \citep{Nelson:2015}, or Illustris TNG \citep{Nelson:2019} typically only marginally, if at all, resolve dwarf galaxies in halos with $\Mv\sim (10^8-10^{10})\;\msun$ where this effect is expected to be most prominent. Such low mass galaxies can be well-resolved in ``zoom-in'' simulations of individual halos in a cosmological context \citep[e.g., FIRE;][]{Fitts:2017,Hopkins:2018}, but this comes at the expense of both small sample sizes of galaxies, as well as much smaller and relatively isolated volumes where the relevant large-scale structures are either missing or significantly under-resolved, since they lie outside the ``zoom-in'' region. In both types of simulation, the resolution is typically adaptive in a quasi-Lagrangian sense such that the characteristic mass element is kept fixed. The spatial resolution thus becomes very poor in the low-density CGM, and even moreso in the IGM \citep[e.g.][]{Nelson:2016}, making it difficult to study the interaction between the gas reservoirs of dwarf galaxies and their intergalactic environment. Properly studying this interaction requires high resolution in both the CGM of dwarf galaxies and the surrounding IGM and cosmic-environment. While several groups have recently developed methods to better resolve the CGM using fixed spatial resolution \citep[e.g.,][]{vandeVoort:2019,Peeples:2019,Hummels:2019}, the refined regions in these simulations typically only extend $\sim (1-2)\; R_{\rm v}$ from the central galaxy of interest. Indeed, such fixed volume resolution frameworks cannot be readily extended to the IGM due to its immense volume.

To bridge this gap, \cite{Mandelker:2019,Mandelker:2021} presented a novel suite of cosmological simulations zooming-in on a large patch of the IGM in between two massive halos which by at $z\sim 2$ are $\Mv\sim 5\times 10^{12}$ M$_{\odot}$ and are connected by a $\sim$ Mpc scale cosmic filament. We hereafter refer to this suite of simulations as \hypa{}, where ``IPM'' refers to the ``intra-pancake medium'' of multiphase gas within cosmological sheets (or ``Zel'dovich pancakes,'' see \citealp{Mandelker:2021}). These are among the highest resolution simulations of such a large region of the IGM to date, with resolution characteristic of a standard ``zoom-in'' simulation of a single halo applied within the entire IGM region between the two protoclusters. One notable feature of this simulated region is the collision of two cosmological sheets between $z\sim5.5$ and $z\sim 4.5$. This collision and subsequent joining of the sheets produces an accretion shock that propagates across the sheet region, rapidly heating much of the IGM gas in the sheet to $T\gsim 10^6$ K --- providing an ideal laboratory for studying the effects of such large scale accretion shocks on dwarf galaxies. A detailed convergence study of the gas properties in this region presented in \cite{Mandelker:2021} suggests that the structure of the IGM in general, and the IPM in particular, is poorly resolved in standard large-box cosmological simulations (e.g., TNG100), and that much like the CGM, is not fully converged even at the resolution of some zoom-in simulations. The resolution in \hypa{}, however, appears sufficient to resolve the formation of multiphase gas in the turbulent IPM, though the morphology and distribution of the cold component is likely only marginally resolved \citep{Mccourt:2018,Mandelker:2019,Mandelker:2021,Gronke:2021}. \hypa{} thus has both the spatial extent to capture relevant numbers of dwarf galaxies in the sheet environment and the resolution to model galaxy-shock interactions.

In this study, we investigate the degree to which the shock-heated sheet acts as a driver for quenching by compiling the star formation histories, baryonic mass accretion histories, bound gas masses in the CGM, and measures of the ambient temperature for galaxies in the \hypa{} simulation which live in the sheet region and experience the accretion shock. By temporally correlating the downturns in star formation, rises in ambient temperature, and decreases in mass accretion rates, we determine for which systems the transition to a shocked sheet environment was responsible for their ultimate quenching. By examining similar regions in TNG50, we estimate how common such systems and quenching mechanisms are likely to be.

In Section \ref{methods}, we present the simulation, the algorithm used to track haloes over time, the sample selection, and the set of simulation properties we extract for each system. In Section \ref{results}, we present the evolution of the selected dwarf galaxies as they interact with the shock-heated sheet. In Section \ref{discussion}, we discuss our results in the context of TNG50, a larger volume simulation, investigating how the quenched fraction in our small volume compares to similar regions across TNG50, and examine how much gas is deposited into the IGM by stripped dwarf galaxies. In Section \ref{summary}, we summarize our results and present conclusions. Throughout, we assume a flat $\Lambda$CDM cosmology with $\Omega_{\rm m}=1-\Omega_{\Lambda}=0.3089$, $\Omega_{\rm b}=0.0486$, $h=0.6774$, $\sigma_8=0.8159$, and $n_{\rm s}=0.9667$ \citep{Planck16}.

\section{Methods}
\label{methods}

A detailed description of the simulation used here is presented in \cite{Mandelker:2019} and \cite{Mandelker:2021}; the reader is directed there for a detailed discussion. A brief summary of the simulation is provided here for reference. 

\subsection{Hydrodynamic Simulation}
\label{sec:Sim}

The \hypa{} simulation was run using the quasi-Lagrangian moving-mesh code \texttt{AREPO} \citep{Springel:2010}. The goal of \hypa{} was to simulate not just a single halo or cluster region as is typically done, but rather two proto-clusters and the intervening space between them where large scale structure evolves. The target haloes were selected by first considering the 200 most massive haloes at $z\sim2.3$ in the Illustris TNG100 magnetohydrodynamic cosmological simulation \citep{Pillepich:2018a,Nelson:2018,Springel:2018}. These haloes span a mass range of $\Mv\sim (0.7-2)\times 10^{12}\;\msun/h$, where $\Mv$ is the virial mass defined using the \cite{Bryan:1998} spherical overdensity. Then, all pairwise combinations of those haloes with comoving distances between 2.5 and 4.0 Mpc/$h$ were identified, comprising 48 pairs, and it was determined that all were either connected by cosmic filaments with radii comparable to the halo virial radius, or else resided in the same cosmic sheet with thickness comparable to the halo virial radius. A single pair was chosen at random to re-simulate as \hypa{}. The pair comprised two haloes with masses of $\Mv\sim$5$\times$10$^{12}$\Msun{} each, separated by a proper distance of $D\sim1.2$ Mpc at $z\sim 2.3$. At $z=0$, these haloes have masses between 1.6 and 1.9 $\times$ 10$^{13}$ \Msun{} and are $\sim$2.7 Mpc apart. 

All dark matter particles in a large region, comprising two spheres centered on the two haloes, with radii equal to the larger of the two virial radii, and a cylinder of the same radius encompasing the filament which connects the haloes, were traced back to the initial conditions of the simulation at $z=127$, where the region was refined to higher resolution (see below) and re-run through $z=2$. The sheet-collision which causes the shock begins at $z\sim$5\, and sweeps across the entire sheet volume by $z\sim 4$ (Figure \ref{fig:fullmaps}). The simulations were performed with the same physics model used in the TNG100 simulation, described in detail in \cite{Weinberger:2017} and \cite{Pillepich:2018b}.

We follow the production and evolution of nine elements (H, He, C, N, O, Ne, Mg, Si, and Fe), produced in supernovae Type Ia and Type II and in AGB stars according to tabulated mass and metal yields. An ionizing ultraviolet background (UVB) from \citet{FG09} is instantaneously switched on at $z=6$ and is assumed to be spatially uniform but redshift dependent. Self-shielding from the UV background is implemented using fits to the degree of self-shielding as a function of hydrogen volume density and redshift in radiative transfer simulations following \citet{Rahmati:2013}. Hydrogen and Helium cooling are included using the analytical formulae of \citet{Katz:1996}, while metal line cooling is included using pre-calculated rates as a function of density, temperature, metalicity and redshift \citep{Wiersma:2009}, with corrections for self-shielding. Cooling is further modulated by the radiation field of nearby active galactic nuclei (AGN) by superimposing the UVB with the AGN radiation field within $3$ \Rv{} of halos containing actively accreting supermassive black holes \citep{Vogelsberger:2013}. Gas with density greater than $n_{\rm thresh}=0.13$ cm$^{-3}$ is considered eligible for star-formation and is placed on an artificial equation of state meant to mimic the unresolved multiphase ISM \citep{Springel:2003}. We include feedback from both supernova and AGN following the Illustris TNG model.

The output of the simulation comprises both particle data and group catalogs. To identify groups and halos, we first apply a Friends-of-Friends (FoF) algorithm with a linking length $b = 0.2$ to the dark matter particles, then assign gas and stars to FoF groups based on their nearest-neighbour dark matter particle. Within each identified group, particles are sorted into substructures (``subhaloes'') via the \texttt{subfind} algorithm \citep{Springel:2001,Dolag:2009}. For the identified groups and \texttt{subfind} objects, catalogs are created which store basic properties of each FoF group and \texttt{subfind} object within the simulation calculated on-the-fly, e.g., total mass and the mass of each component (gas, stars, and dark matter), star formation rate, and gas metallicity. We denote in text whenever we have used a quantity present in the group catalogs for a given object, or whether we compute it directly from the particle data. We treat the 0th \texttt{subfind} subhalo of each group as the central (which generally contains $>90\%$ of the group mass). All galaxies presented in this work are centrals. In general, properties of the FoF group (total mass, SFR, etc.) and properties of the 0th subhalo identified by \texttt{subfind} are consistent with one another. During analysis, we carried out calculations with both to ensure consistency, and in this work present results generally calculated using the \texttt{subfind} value unless otherwise noted. Certain properties (e.g.,\Rv{}) are only defined in practice for the groups, for example.

Tracking the properties of our identified sample of haloes over cosmic time in our simulation necessitates the construction of merger trees. In this work, we use an updated version of the linking algorithm first presented in \cite{Springel:2001}, and detailed in \cite{Springel:2021}, to track haloes forward and backward in time. The primary update to the algorithm concerns the determination of the ``primary'' which constitutes the main branch of the progenitor/descendant tree. Previously, the most massive object was selected automatically --- however, this becomes ambiguous when two similarly massive objects merge and their mass difference is within the noise limit. In this updated approach, a recursive process is applied which determines which halo contributes more total mass over the rest of the simulation.

The simulation was run at multiple resolutions, with a detailed convergence study of the gas properties in the IGM (in both filaments and the sheet) presented in \cite{Mandelker:2021}. In this work, we use zoom factor 3, (hereafter ZF3), the second-highest resolution version (see \citealp{Mandelker:2021}). This version has a dark matter particle mass of $m_{\mathrm{dm}}=1.9 \times 10^{5} \;M_{\odot}$ and a Plummer-equivalent gravitational softening of $\epsilon_{\mathrm{dm}}=333 \mathrm{pc}$ comoving. Gas cells are refined such that their mass is within a factor of 2 of $m_{\mathrm{gas}}=3.5 \times 10^{4} \;M_{\odot}$ and have a minimal gravitational softening $\epsilon_{\mathrm{gas}}=0.5 \epsilon_{\mathrm{dm}}$. This mass resolution is a factor of $\sim$27 higher than TNG100, and a factor of $\sim$3.3 higher than TNG50. We selected this resolution as the highest resolution run, ZF4, was terminated at $z\sim 3$, whilst our analysis extends to $z\sim 2$.

For convenience, in this work we will refer to several commonly-discussed snapshots as follows: Snap($z=5.519$) as $z=5.5$, Snap($z=5.078$) as $z=5$, Snap($z=4.055$) as $z=4$, Snap($z=2.99$) as $z=3$, and Snap($z=2.154$) as $z=2.2$.

\begin{figure}
    \centering
    \includegraphics[width=\linewidth]{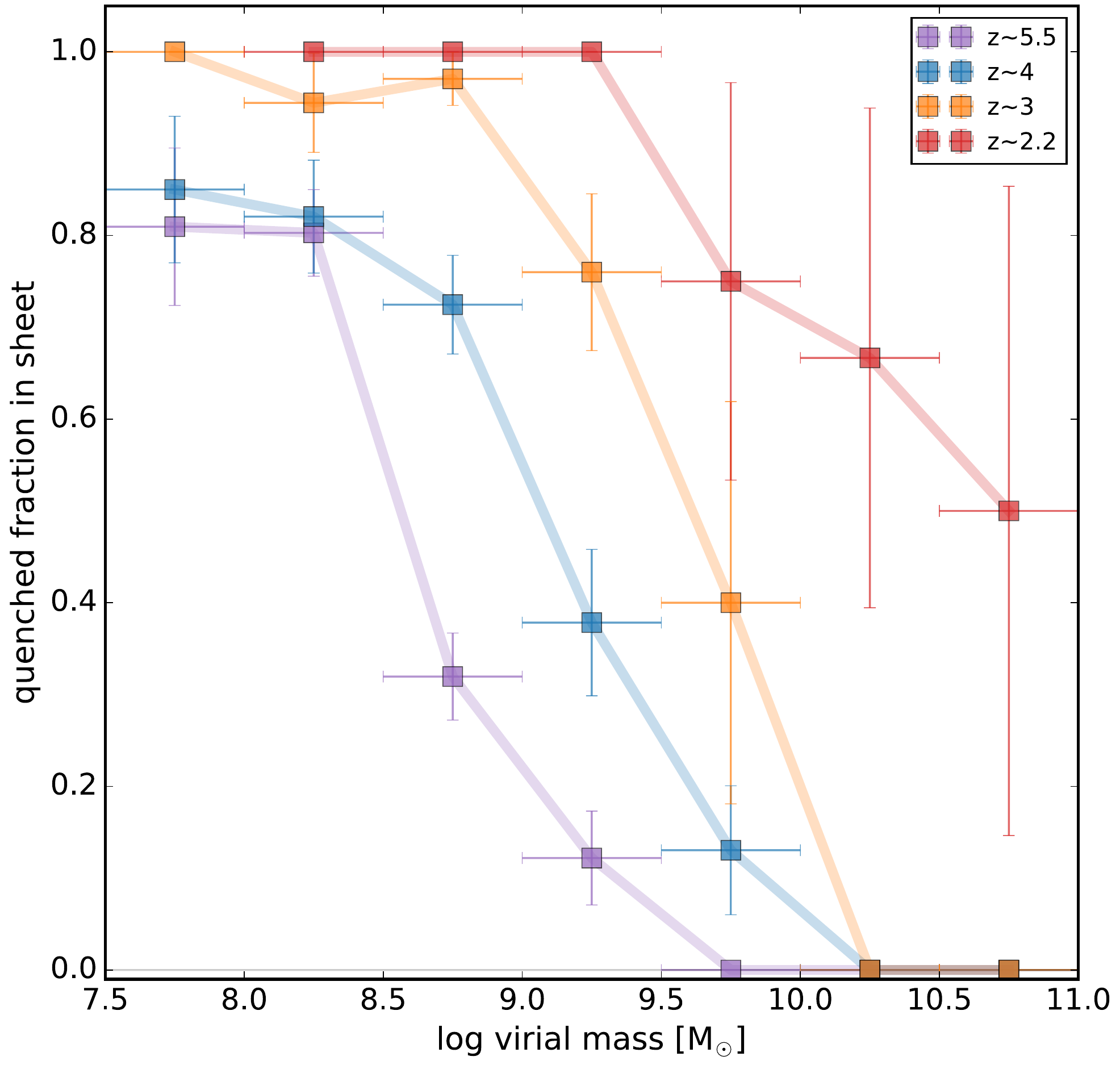}
    \caption{Quenched fraction for all galaxies in the \hypa{} simulation which are in the sheet region (whose thickness is estimated as the virial radius of the proto-cluster at each redshift), but are more than 3 \Rv{} away from the three most massive haloes, for the four redshifts shown in Figure \ref{fig:fullmaps}. Horizontal error bars show the bin sizes used, while the uncertainty in quenched fraction is estimated using the standard error of percentages, $\sqrt{f_Q(1.0-f_Q) / N_{\rm gal}}$. In the $z=2.2$ snapshot, there were no galaxies meeting the selection criteria in the $7.5 < M_{\rm V}<8.0$ bin. Uncertainty in $f_Q$ at this snapshot is also large because the \Rv{} of the two proto-clusters are so large that little of the sheet remains beyond 3 \Rv{}. Systems with $M_*=0$ were excluded when calculating the quenched fraction. Particularly in the mass range $8.5<\log M <10.0$, the quenched fraction is a strong function of redshift, with its value increasing as time progresses and the shock fills the sheet volume. Note that this is the mass regime for which in the local universe, field dwarf galaxies have a quenched fraction of $\sim0$. }
    \label{fig:qf}
\end{figure}

\subsection{Sample Selection}
\label{sec:Selec}

We select our sample of dwarf galaxies in the $z=3$ snapshot of the simulation, taking all galaxies which have 
\begin{itemize}
\item Accrued virial masses such that $10^{8.5}$ M$_{\odot}$ $<$ $\Mv$ $<$ $10^{10.5}$ M$_{\odot}$, 
\item Ultimately (by $z=2$) form at least $10^{5.5}$ M$_{\odot}$ in stars,
\item That are, in the $z=3$ snapshot, at least 3 \Rv{} away from the three most massive haloes (which eventually form the proto-clusters at z$\sim2$),
\item That do not become satellites of other systems during the simulation before quenching (see below), and, 
\item Which are spatially located in the sheet region that spans the region between the massive proto-clusters in the $z\simeq3$ snapshot.
\end{itemize}
Assessment of co-spatiality with the sheet was determined visually using two orthogonal projections (face-on and edge-on) of the sheet region's temperature and density distribution; the sheet structure is well-defined, at least until $z\sim3$, with temperature and density dropping sharply at the sheet boundaries. Figure \ref{fig:fullmaps} shows projection maps of the density-weighted gas temperature projected along the line-of-sight both face-on (left) and edge-on (right) with respect to the sheet, at redshifts of 5.5, 4, 3 and 2.5 (from top to bottom).  At the selection redshift ($z=3$), all of our galaxies live within the shock-heated sheet (see, i.e., right panels of Fig. \ref{fig:fullmaps}). We select the chosen mass range to target galaxies whose virial temperatures (Eqn. 1) are near the temperature of the ambient  gas in the sheet region prior to the shock. The stellar mass cutoff was selected both to avoid low-count statistics near the resolution limit of the simulation, as well as to target primarily systems that have formed enough stars to be relevant observationally --- e.g., Crater II with $M_{*}\; \sim2.56\times10^5$ M$_{\odot}$ \citep[][]{Torrealba:2016,borukhovetskaya:2021}. 

The condition on distance from the three most massive haloes is to ensure that any environmental quenching we may find is not due to the influence of an individual halo. Within the simulation, cooling is modulated by the radiation field of nearby active galactic nuclei (AGN), via the superposition of the global UV background and the radiation field within 3 $R_{\rm V}$ of haloes containing AGN \citep{Vogelsberger:2013}, so imposing a $d>$ 3 $R_{\rm V}$ condition ensures the ambient temperature in the environment of a selected galaxy is set by the large-scale shock, not AGN. 

As a note, we experimented with selecting additional samples at earlier and later epochs. Galaxies which meet the above requirements at earlier redshifts (e.g., $z=5.5$) generally fall into one of the massive proto-clusters significantly before $z\sim2$, as they spend more time in the region near these haloes. Alternatively, defining the sheet becomes progressively more difficult after $z\sim3$ as it is in the process of further collapsing into a filament, and galaxies which enter the region this late in the simulation run the risk of containing low resolution particles, and do not have as much time evolution within the sheet. 

The above cuts produce a sample of \Ngal{} galaxies. 40 met the mass and position cut; 8 of those were found to be satellites at earlier times and were removed. In one unique case among the sample, two galaxies of similar mass (173 and 194, see Table \ref{tab:sample}) eventually come to ``share'' the same halo. The merger tree designated 194 as a satellite of 173 in this case, as it is slightly less massive. We leave these systems in as the merger occurs after both are embedded in the hot sheet and are of similar mass (i.e., one is not quenching the other due to its hot halo), but our results are unchanged if these systems are excluded. Some of the galaxies in the final sample are temporarily (incorrectly) \textit{classified} by the merger trees as satellites; we discuss the handling of these cases explicitly in Section \ref{corrections}. 

\subsection{Quenching Definition}
There are several ways to define a system as quenched. Here, we adopt a quenching threshold of 1 dex below the redshift dependent star forming main sequence (SFMS) presented in \cite{Stark:2013}. We interpolate the redshift dependence of the normalization of the SFMS using a fourth order polynomial between $7.0<z<2.0$ given by
\begin{equation}
    \log sSFR(z)_{\rm MS} = a_0 + a_1z + a_2z^2 + a_3z^3 + a_4z^4
\end{equation}
with coefficients
\begin{align*}
a_0 &= -9.004  & a_1 &= 0.077  & a_2 &= -0.46  & \\
a_3 &= 1.27 & a_4 &= -0.83.
\end{align*}
This relation is consistent with other studies \citep[e.g.,][]{Davidzon:2017}, and the mass range discussed in this work is in the regime for which the sSFR is measured to be roughly independent of stellar mass \citep[e.g.,][]{Popesso:2022}.
Many systems in our sample also ultimately meet the common observational threshold of log sSFR $< -11.0$, and indeed, drop to SFRs of 0 in the simulation. We classify a system as quenched by the sheet if it both drops below the threshold, and has its final snapshot (either before the end of the simulation or before moving within 3 \Rv{} of a massive halo) below the threshold, after interacting with the shock. In Table \ref{tab:sample}, we show the full sample, bolding Galaxy IDs for systems which meet this quenching definition.  A few of the close cases (or systems which oscillate around the quenching threshold but end above it) are discussed in the comments column of the table, as well as several systems which drop close to, but do not quite reach, our threshold. We discuss these cases specifically in Section \ref{results}.

\subsection{Galaxy Properties}
We track each of the galaxies selected at $z=3$ using the criteria described in Section \ref{sec:Selec} forward and backward in time, using the merger trees described in Section \ref{sec:Sim}, and calculate the following quantities of interest for each galaxy in our sample: 
\begin{itemize}
    \item \textbf{Total Bound Gas Mass}. Stored in the simulation catalogs, this property constitutes the mass of all gas cells considered gravitationally bound to the halo according to \texttt{subfind}. Other catalog-tracked properties, such as the bound dark matter mass, and bound stellar mass, are also extracted.  
    \item \textbf{ISM and CGM gas.} We divide the total bound gas mass of each galaxy into ISM- and CGM-like phases. As these properties are not listed in the group catalogs, we compute them in post-processing using the particle data. For each gas cell which is bound to a galaxy according to \texttt{Subfind}, we evaluate the hydrogen number density, $n_{\rm H}$, and define this cell as being in the ISM or CGM of the galaxy to which it is bound if $n_{\rm H}>0.1\,{\rm cm}^{-3}$ or $n_{\rm H}<0.1\,{\rm cm}^{-3}$, respectively. This is approximately equal to the threshold density for star-formation used in the simulation (see Section \ref{sec:Sim}), above which the gas cell is placed on an artificial equation of state meant to mimic the multiphase ISM \citep{Springel:2003}. We crudely associate this density threshold with a temperature threshold, as star-forming gas on the artificial equation of state does not have a meaningful temperature associated with it.
    \item \textbf{Extant Gas Mass.} Using the particle data, we also calculate the mass of all gas cells within \Rv{} of each halo, whether bound or unbound. This is useful for determining whether drops in bound gas mass are due to gas being removed from the vicinity of the halo or whether gas is still present, but is no longer bound, either because it has been heated to temperatures $T>T_{\rm v}$, because it is being stripped, or because it has been expelled via strong feedback.
    \item \textbf{Ambient Temperature.} We calculate the ambient mass-weighted temperature in a shell extending from $1$--$2$ \Rv{}. This quantity is compared to the virial temperature of the halo to assess its ability to retain its bound gas.
    \item \textbf{Stellar Mass.} We extract the bound stellar mass for each system directly from the catalogs.
    \item \textbf{Star Formation Rate.} From the simulation group catalogs, we extract the star formation rate of each halo, and additionally compute the specific star formation rate (sSFR $=SFR/M_*$) using the stellar mass (above).
    \item \textbf{Gas Accretion}. We calculate the inward mass accretion rate of gas (in M$_{\odot}$ yr$^{-1}$) by selecting gas cells in a shell extending from
    $R_{\rm min}=R_{\rm v}$ to $R_{\rm max}=2R_{\rm v}$ whose net radial velocity is negative (inward) and which have a radial velocity magnitude greater than half the virial velocity of the halo. Using these cells, we calculate the mass-weighted inward flux as
    \begin{equation}
        \left(\frac{\Delta M}{\Delta t}\right)_{\rm in} = \frac{\sum m_i |v_{r,i}|}{R_{\rm max} - R_{\rm min}},
    \end{equation}
\end{itemize}

\begin{figure*}
    \centering
    \includegraphics[width=\textwidth]{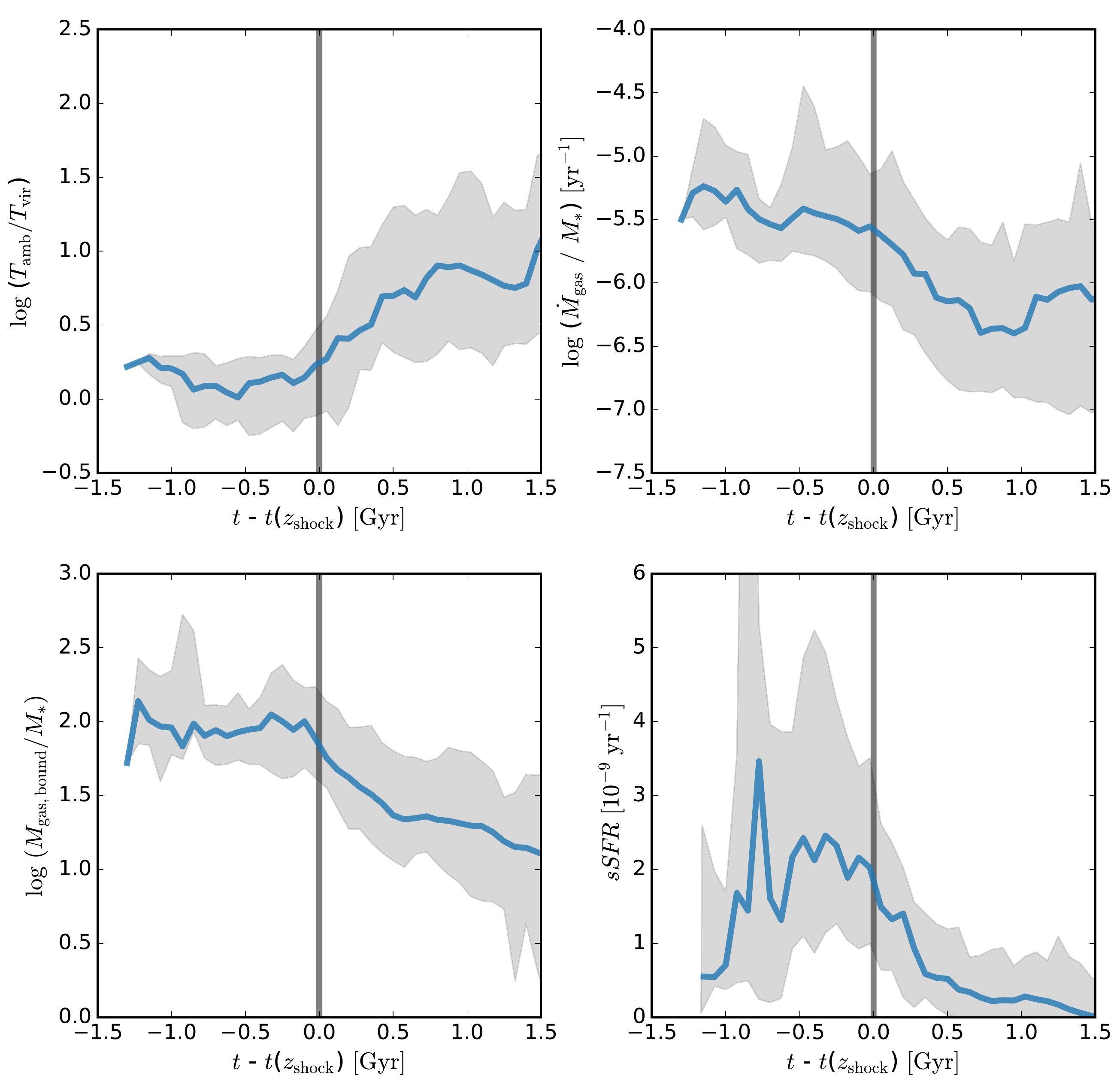}
    \caption{Properties of the full sample, normalized by the time at which each first interacted with the shock-heated sheet (generally between $z=5$ and $z=3$). Across all galaxies, we compute the mean sample behavior, as well as the 16th and 84th percentiles. Note that as many systems drop to having 0 sSFR, we show the sSSFR comparison of panel 4 in linear units. We find that across the galaxies in the sample, there is a strong trend toward increases in $T_{\rm amb} / T_{\rm V}$ just after entering the shocked region, mirrored by decreases in bound gas fractions and sSFR. Gas accretion onto the haloes in the sample appears to also decrease after shock interaction, but unlike the other properties, increases again for many of the systems after $\sim$1.5-2 Gyr. The mean quenching time for galaxies in the sample is short, as sSFR tends to drop rapidly at the shock onset. We find that $\langle t_{\rm quench}\rangle = 275$ Myr, with $\langle t_{\rm quench}\rangle$ defined as the time for the mean sSFR relation to drop by 1/2 after $t_{\rm shock}$.}
    \label{mean_behavior}
\end{figure*}

\begin{figure*}
    \centering
    \includegraphics[width=\textwidth]{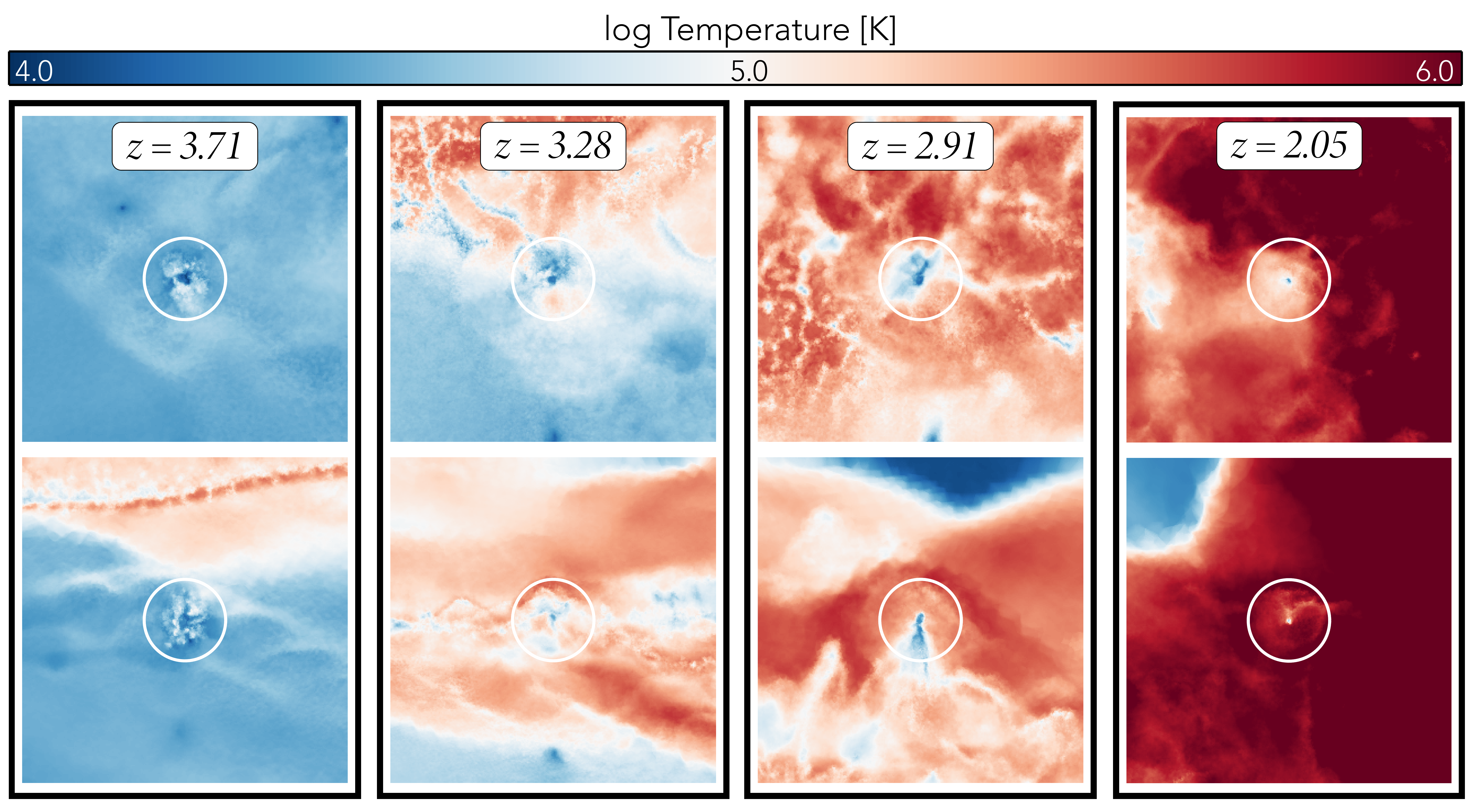}
    \caption{Density-weighted temperature maps of Galaxy 61 ($M_*\simeq10^7 M_{\odot}$ at $z\sim 2$) from face on (top) and edge on (bottom) viewing angles with respect to the sheet, at four redshifts. The integration depth is set to 10 \Rv{}, and $R_{\rm V}$ in each snapshot is denoted with a white circle. This system enters the region after the establishment of the shock, and once it does, begins losing cold gas, ultimately quenching. In the third panel, cool gas is being stripped from the system (presumably via ram pressure stripping).}
    \label{fig:g61_temporal}
\end{figure*}

\begin{table*}
\begin{tabular}{ccccccl}\hline
\multicolumn{1}{c}{Galaxy ID} & \multicolumn{1}{c}{log $M_{*,f}$} & \multicolumn{1}{c}{log $M_{\rm gas,bound}$} & \multicolumn{1}{c}{$M_{\rm gas,f} / M_{\rm gas,max}$} & \multicolumn{1}{c}{$z_{\rm shock}$} & \multicolumn{1}{c}{sSFR$_{f}$ / sSFR$_{z_{\rm shock}}$} & Comments \\
\multicolumn{1}{c}{} & \multicolumn{1}{c}{($z\simeq 2$)} & \multicolumn{1}{c}{(maximum)} & \multicolumn{1}{c}{} & \multicolumn{1}{c}{(first interaction)} & \multicolumn{1}{c}{} &  \\\hline
17 & 8.3 & 9.5 & 0.83 & 4.4  & 0.24 & \\
18 & 8.5 & 9.4 & 1.00 & 4.8  & 0.25 & \\
28 & 7.8 & 9.2 & 0.89 & 3.7 & 0.36 & Temporarily shielded in filament \\
32 & 7.5 & 8.9 & 0.93 & 3.5 & 0.49 & \\
35 & 7.0 & 8.5 & 0.57 & 3.2 & 0.39 & Falls with 3 $R_{\rm V}$ of cluster just after $z=3.0$\\
36 & 7.5 & 9.0 & 1.00  & 3.6 & 0.31 & Falls with 3 $R_{\rm V}$ of cluster just after $z=2.5$\\
45 & 7.7 & 8.9 & 0.57 & 4.0  & 0.19  & Rapidly approaching quenching threshold \\
\textbf{61} & 7.1 & 8.5 & 0.20 & 3.3 & 0.10 &  \\
\textbf{69} & 6.3 & 8.2 & 0.54 & 3.8 & 0.06 &  \\
\textbf{76} & 7.0 & 8.5 & 0.15 & 4.5 & 0.10 &  \\
\textbf{88} & 6.9 & 8.4 & 0.03 & 4.2 & 0.00 &  \\
92 & 6.3 & 8.3 & 0.48 & 4.2 & 0.09 & Quenched in 9 of final 20 snapshots \\
\textbf{95} & 6.0 & 8.0 & 0.34 & 4.5 & 0.03 &  \\
\textbf{99} & 5.6 & 7.8 & 0.00 & 4.7 & 0.00 &  \\
\textbf{103} & 6.5 & 8.2 & 0.04 & 5.0 & 0.002  &  \\
\textbf{107} & 6.4 & 8.0 & 0.53 & 3.8 & 0.09 & Falls with 3 $R_{\rm V}$ of cluster just after $z=2.9$\\
115 & 6.5 & 8.1 & 0.47 & 4.0 & 0.21  & Quenched in 8 of final 20 snapshots \\
\textbf{117}& 5.5 & 7.5 & 0.00 & 3.7  & 0.00 &  \\
\textbf{124} & 6.1 & 8.0 & 0.00 & 4.5 & 0.00 &  \\
\textbf{125} & 6.0 & 8.1 & 0.39 & 3.3 & 0.12 & \\
\textbf{127} & 5.9 & 8.4 & 0.04 & 4.4 & 0.00 & Spends time in a filament  \\
\textbf{128} & 5.5 & 7.6 & 0.00 & 4.2 & 0.00 & Becomes satellite of 69 after quenching \\
\textbf{136} & 5.7 & 7.7 & 0.18 & 3.9 & 0.01 &  \\
140 & 5.7 & 7.8 & 0.79 & 4.0 & 0.16 & \\
\textbf{151} & 5.5 & 7.8 & 0.00 & 4.8 & 0.00 &  \\
\textbf{173} & 5.9 & 7.7 & 0.00 & 4.0 & 0.00 & co-primary with 194 after  $z\sim3$ \\
\textbf{190} & 5.5 & 7.3 & 0.00 & 3.0  & 0.00 &  \\
\textbf{192} & 6.1 & 8.0 & 0.00 & 4.7 & 0.00 & briefly a satellite of 115 \\
\textbf{194} & 5.7 & 7.6 & 0.00 & 4.1 & 0.00 & co-primary with 173 after $z\sim3$ \\
\textbf{195} & 5.6 & 7.6 & 0.00 & 4.7 & 0.00 &  \\
\textbf{210} & 5.9 & 7.8 & 0.00 & 5.0 & 0.00 &  \\
\textbf{251} & 5.7 & 7.8 & 0.00 & 4.8  & 0.00 &  \\\hline
\end{tabular}
\caption{Sample of galaxies selected in the sheet region at $z=3$. Final ($z=2$) stellar masses, maximum bound gas content, fraction of final gas content to maximum gas content, shock interaction redshift, and ratios of final sSFR to sSFR at the redshift of first shock interaction are noted. In cases where the system falls within 3 \Rv{} of a massive halo after $z=3$, we compute $M_{\rm {gas},f}$ and sSFR$_{f}$ using the last snapshot prior to entering this radius. Galaxy IDs are in bold for systems whose final sSFR (either $z=2$ or last snapshot prior to entering 3 \Rv{} of a protocluster) strictly meets our quenching condition. Galaxy 45 has a final sSFR above the cutoff but is rapidly approaching it. Galaxies 92 and 115 have final snapshots above the cutoff, but near the end of the simulation are oscillating about it ($\sim$half of the last 20 snapshots are below the quenching threshold). Galaxies at the massive end of the sample, despite not quenching, do generally respond to the shock, dropping in sSFR to $\sim$25-50\% of pre-shock levels.}
\label{tab:sample}
\end{table*}

\begin{figure*}
    \centering
    \includegraphics[width=\textwidth]{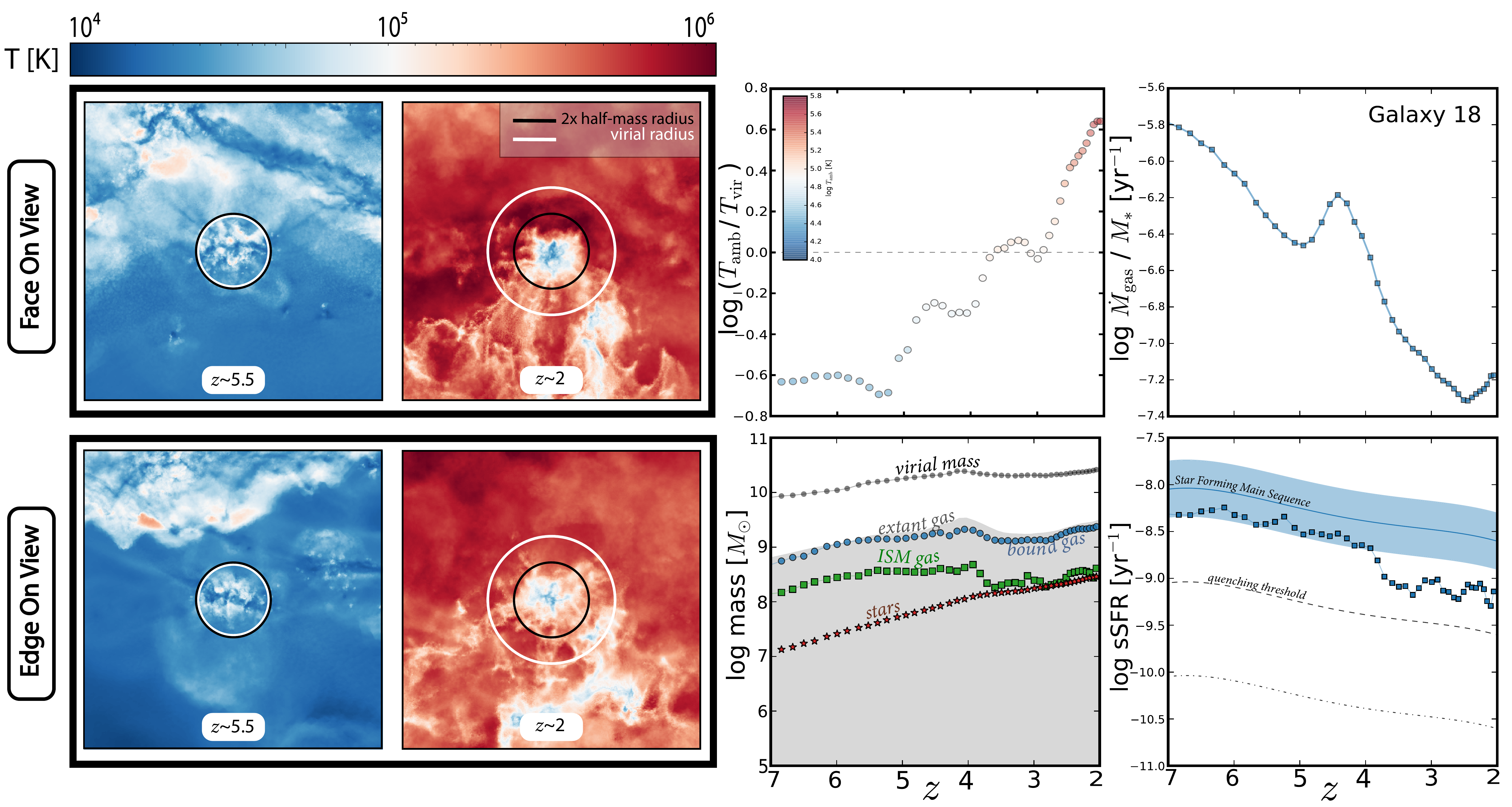}
    \caption{\textbf{\textit{Left}}: Density weighted temperature maps of galaxy 18 both face on (top) and edge on (bottom) with respect to the sheet, at two redshifts selected before and after the shock interaction. The integration depth in each map is 10 $R_V$. \textbf{\textit{Right}}: Four panels displaying properties of galaxy 18 as a function of redshift. In the upper left panel, the ratio of ambient temperature to virial temperature starts strongly in favor of the galaxy, but as the ambient temperature rises, moves above unity. As the gas accretion of the galaxy continues to slow, the ratio drops even further. This can be seen in the upper right panel as well; the specific gas mass accretion rate declines throughout the simulation, with the exception of an enhancement at z$\sim$4 associated with the shock arrival. At this same redshift, the final two panels show the effects on the galaxy; the amount of bound gas being accumulated flattens (lower left panel), and though it continues to grow in stellar mass, the sSFR drops by $\sim0.5$ dex from the previous value (lower right panel), which was until this point within typical scatter for the star-forming main sequence.}
    \label{fig:g18}
\end{figure*}
\begin{figure*}
    \centering
    \includegraphics[width=\textwidth]{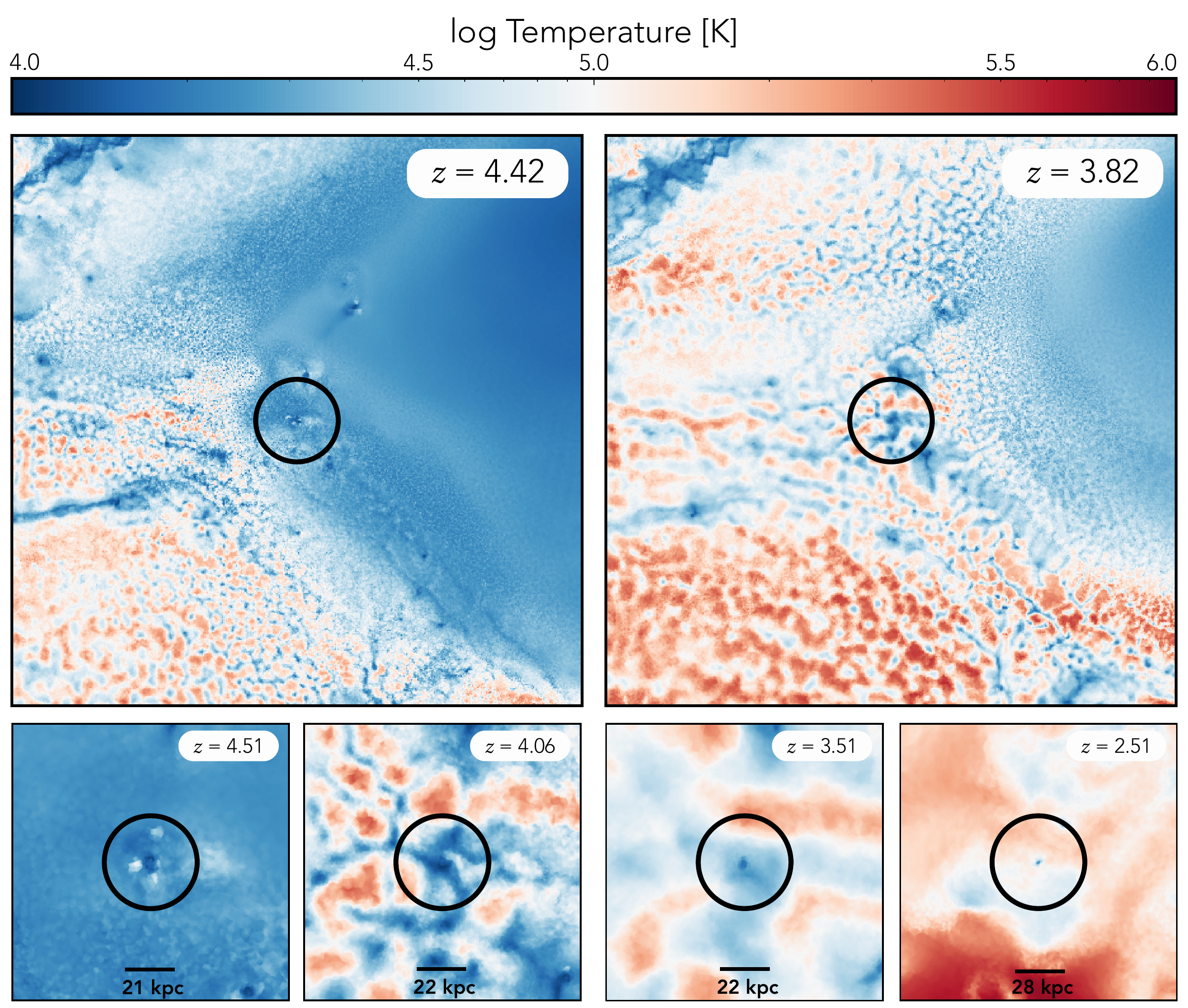}
    \vspace{-10pt}
    \caption{\textit{Top:} Density-weighted temperature maps shown face-on with respect to the sheet at $z\sim4.4$ and $z\sim 4$ for a subregion of the sheet 900 by 900 ckpc/$h$ in size and with an integration depth of 400 ckpc/$h$. Of particular interest is galaxy 173 (circled): the progression of the sheet collision shock can be tracked as it moves rightward in the simulation, and this system gets hit by this shock just before $z=4$. \textit{Bottom:} Zoomed in panels of this system at four redshifts, illustrating how after the arrival of the shock, the region becomes progressively more dominated by hot ($T\geq 10^{5.5}$ K) gas. In all panels, the black circle represents the virial radius of the system.}
    \label{fig:173_shock}
\end{figure*}
In all cases where we refer to the virial radius \Rv{}, we use the \cite{Bryan:1998} spherical overdensity. For quantities determined via analysis of a shell surrounding each galaxy (e.g., mass accretion, ambient temperature), we repeated our analyses using varying shell widths (0.5 $R_{\rm V}$, 1.0 $R_{\rm V}$, and 1.5 $R_{\rm V}$) and shell centers (1.5 $R_{\rm V}$, 2.0 $R_{\rm V}$, and 2.5 $R_{\rm V}$), and confirmed that these choices did not significantly impact any of the trends seen in this work. 

\begin{figure*}
    \centering
    \includegraphics[width=\linewidth]{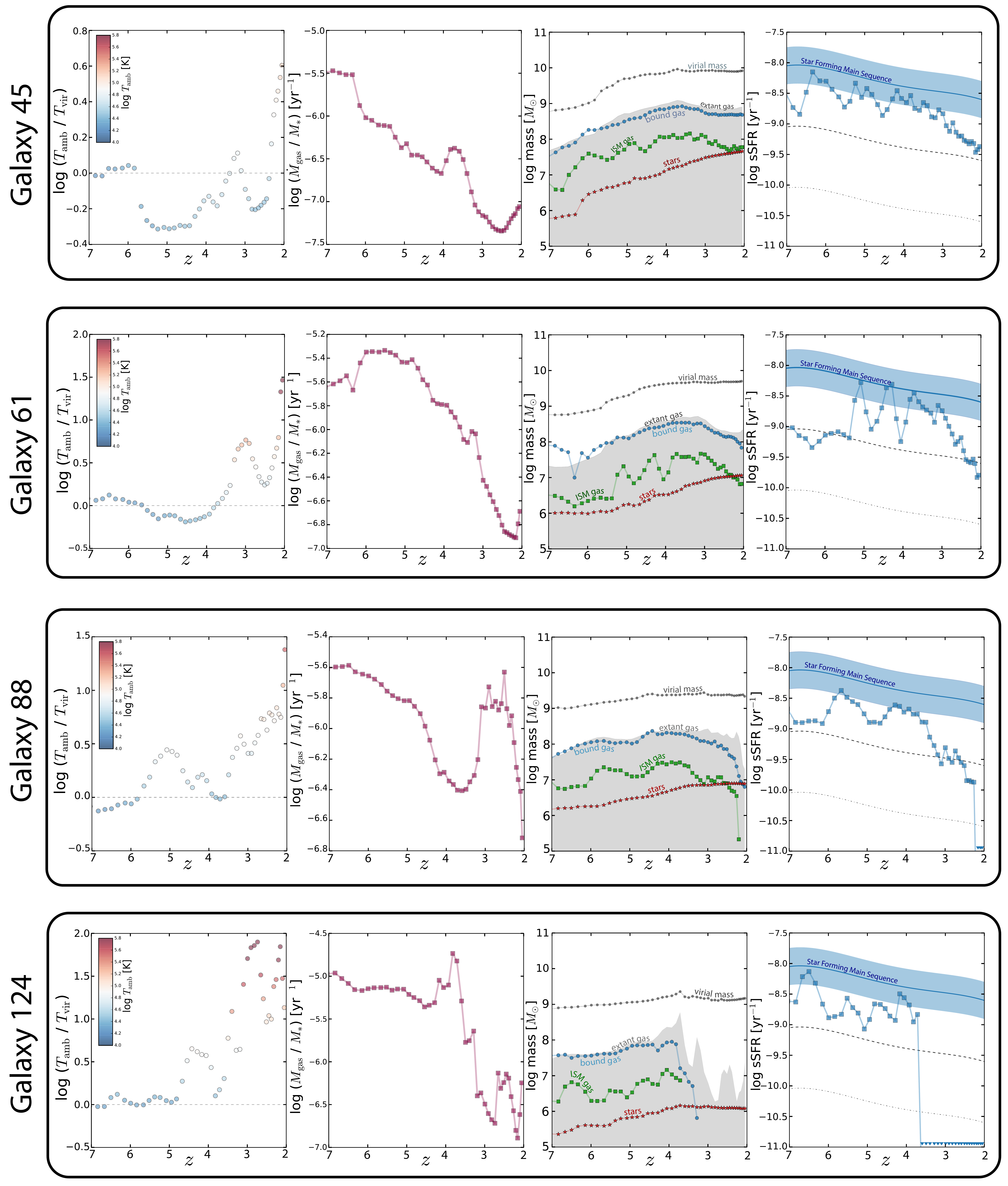}
    \caption{Evolution of galaxies 45, 61, 88. and 124. For all figures of this form, the first panel shows the ratio of ambient to virial gas temperature, the second panel the specific gas accretion rate onto the halo, the third panel the stellar, virial, gas, and stellar masses over time, and the fourth panel the star formation history of selected galaxies in the simulation, with the star forming main sequence of \citet{Stark:2013} demarcated. When galaxies begin interacting with the shocked environment of the sheet, we observe several correlated changes, including rising $T_{\rm amb}$, general reduction of gas accretion onto the halo (with an occasional enhancement when the galaxy and shock first collide), stagnation (or decline) in bound gas growth, and reduction in sSFR, often by a dex or more.}
    \label{quenched_sample}
\end{figure*}

\begin{figure}
    \centering
    \includegraphics[width=0.95\linewidth]{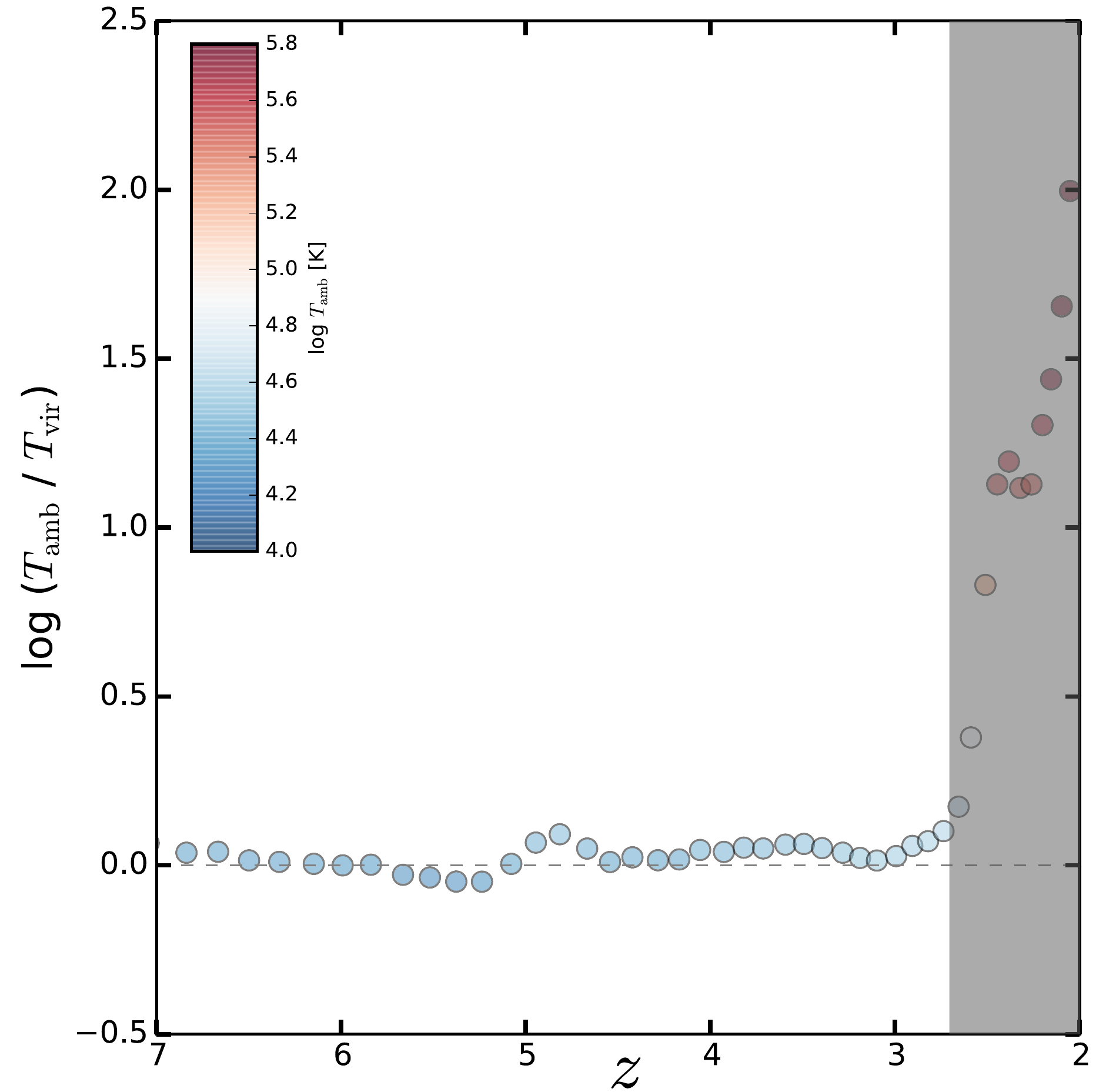}
    \caption{Example of a galaxy which evolves outside the shocked sheet for the majority of the simulation and falls into the cluster at $z\sim 2.5$. The $T_{\rm amb} / T_{\rm V}$ of this galaxy remains steady at unity for the duration of the galaxy's evolution outside the sheet, reflecting the temperature stability in this region compared with the interior of the sheet.}
    \label{outofsheet}
\end{figure}
\subsection{Corrections to Merger Tree Tracking}\label{corrections}

As described above, a subtlety of the merger tree tracking is that individual \texttt{Subfind} objects are tracked, but virial properties ($M_{\rm V}$ and $R_{\rm V}$) are stored in the FoF catalogs. This can lead to issues when, on occasion, spurious dark matter particle bridges between haloes causes the algorithm to assign a given tracked \texttt{Subfind} system to a separate FoF group as a satellite. This sometimes occurs even when the system in question is well beyond the other FoF group's virial radius, and generally occurs only for a single snapshot or two. When this happens, the FoF group virial properties of the tracked object become that of the group it has been linked to, and all computed properties based on the virial radius and mass thus are those of the other group. 

For galaxies in our sample which have this ``glitch'' occur at isolated snapshots, we visually inspect using cutouts the motion of the galaxy and its location with respect to the FoF group to which it is temporarily assigned. In cases where the galaxy is indeed not a satellite, we correct these snapshots by interpolating in log space over $R_{\rm V}$ and $M_{V}$ and recompute the particle data properties now centered on the \texttt{subfind} object using the inferred $R_{\rm V}$ and $M_{V}$.

With corrected histories of the galaxy sample over the length of the simulated constructed, we turn to examining the effect of the accretion shock event on the properties of the galaxies living in the sheet region.

\section{Results}\label{results}

Within the final sample of \Ngal{} galaxies, we find that nearly all exhibit a qualitative response to the accretion shock, generally associated with a drop in both their gas accretion and SFR. Over half of the sample quenches. A majority of these quenching systems (particularly at lower mass) drop to SFRs of 0 before the end of the simulation. 

We find that the primary driver for predicting whether a system in the sheet region quenches or survives the sheet shock is its virial mass accumulated by $z\sim5$, when the shock emerges and starts to propagate the sheet region. Most (but not all) systems that have reached $M_{\rm V}\sim10^{9.5-10}$ M$_{\odot}$ by this point are generally inoculated from the environmental shift, while those with mass less $\lesssim 10^9 M_{\odot}$ are more susceptible to quenching. This overview can be seen in Figure \ref{fig:qf}, which shows the quenched fraction of centrals in the sheet (but $> 3 R_{\rm V}$ away from the massive proto-clusters) as a function of virial mass and redshift. Figure \ref{fig:qf} includes \textit{all} galaxies which at the shown redshift are in the sheet and 3 \Rv{} away from the protoclusters; the \Ngal{} sample represents a slice in mass from this space, with the additional criteria described in Section \ref{sec:Selec}.

Until $z\sim2$, when the temperature of the entire region becomes primarily dominated by the massive proto-clusters, the quenched fraction among galaxies with $M_{\rm V}\gtrsim 10^{10}\msun$ is negligble. On the other hand, galaxies in the range of $M_{\rm V}\sim 10^{8.5-9.5}$ are susceptible to environmental quenching; in particular, the 8.5 $<$ log $M_{\rm V}$ $<$ 9.0 bin exhibits a particularly large increase in $f_Q$ between $z\sim5.5$ and $z\sim4$ --- the interval over which the shock traverses the sheet region. As a note, the lowest mass bins (which we do not consider in this work but which are shown in Figure \ref{fig:qf}) have high $f_Q$ independent of redshift, most likely due to early quenching by reionization. We emphasize that Figure \ref{fig:qf} does not explicitly account for potential backsplash galaxies in any given mass bin, so such systems may contaminate the quenched fraction. This is not the case for the \Ngal{}  sample, for which centrals that were previously satellites were removed.

\subsection{Mean Sample Behavior}
Here we stack the galaxies in the \Ngal{} sample to estimate the mean response of galaxies in this mass range to the changing environment in the sheet region (Figure \ref{mean_behavior}). To do so, we shifted each of the galaxies in the sample onto a common time axis normalized by the time of the shock's arrival (determined by visual inspection of the galaxy stamp movies). Once each galaxy was shifted, we computed the mean behavior of the sample, along with the 16th and 84th percentiles, for four diagnostic properties: (1) the ratio $T_{\rm amb} / T_{\rm V}$, (2) the specific gas mass accretion rate $\dot{M}$ / $M_{*}$, (3) bound gas mass (normalized by stellar mass), and (4) sSFR. Due to the number of galaxies for which the sSFR drops to zero we show the fourth panel in linear scale.

We find that all four of these properties have distinct breaks in their evolution almost immediately upon interacting with the shock (whether by entering the sheet or, from within the sheet, experiencing the arrival of the shock). The sharp rise in ambient temperature, compared to each galaxy's $T_{\rm V}$, is seen in the upper-left panel, and the subsequent declines in bound gas mass and specific star formation rate are evident in the bottom panels. We find a mean quenching time $\langle t_{\rm quench}\rangle=275$ Myr (defined as the time for sSFR to drop by 1/2 after $t_{\rm shock}$). Figure \ref{mean_behavior} shows that the sSFR drops quickly after shock interaction before flattening out (though still declining) between 0.5 and 1.5 Gyr post-shock arrival. 

Figure \ref{mean_behavior} demonstrates that large scale accretion shocks, like the volume-filling shock in the \hypa{} cosmic sheet, can have a sudden and profound effect on dwarf galaxies. Even systems which do not quench as a result (as is the case for some of the massive-end galaxies in our sample) see their bound gas fractions dropping and star formation declining by significant fractions of their pre-shock values. The less massive dwarf galaxies, meanwhile, tend to lose \textit{all} their bound gas, and stop forming stars \textit{entirely}. 

These results are in line with the prediction that cosmic-scale environments can, under certain circumstances, be responsible for quenching central galaxies or affecting their evolution in ways similar to that seen for satellites in galaxy groups and clusters at low redshift. We discuss the implications of this for pre-processing as a method of quenching dwarf galaxies in isolation in Section \ref{discussion}. Before this, however, we split our sample into several subsamples of interest, and examine more closely the impact of this environmental effect on those galaxies.

\subsection{Effect on Massive Dwarfs}

The galaxies in the final sample which have accumulated the highest virial and stellar masses show the widest range in response to the arrival of the accretion shock. If we examine only those systems in the sample with $M_{*,z=2}>10^7$ M$_{\odot}$ ($N$=9), four fully or nearly fully quench (45, 61, 76, 88), an additional system drops by more than 0.3 dex in sSFR and remains sub main-sequence for the remainder of the simulation (18), and the rest (17, 28, 32, and 36) generally exhibit a temporary drop in sSFR before rejoining the main sequence or otherwise show minor responses to the shock. 
 
Examination of each galaxy and its immediate surroundings (within $\sim4$ $R_{\rm V}$) over time\footnote{Movies of each galaxy are available at http://ipm-dwarfs.github.io.} suggest that a combination of general starvation (hot gas medium $>>T_{\rm V}$) and ram pressure stripping as the shock hits the galaxies (and their subsequent movement within the shock-heated sheet) are responsible for the ultimate quenching of the subset that do quench. An illustrative example is provided in Figure \ref{fig:g61_temporal}, which shows the temporal evolution of galaxy 61, a dwarf which by the end of the simulation ($z\sim$2) has a virial mass of $\sim10^{9.7}$ M$_{\odot}$ and stellar mass of $\sim10^{7}$ M$_{\odot}$. In the four provided snapshots, the galaxy can be seen fully entering the shocked region of the sheet at $z\sim3.3$; in the $z= 2.9$ snapshot, a clear plume of cool gas can be seen being stripped off of the galaxy. By $z\sim2$, the majority of the cold gas associated with the galaxy has either been stripped or heated to $T>10^{5-6}$ K. Galaxy 61 retains only 20\% of the bound gas it possessed at the point it entered the shock, and drops in sSFR to the quenching threshold. 

In the case of these most massive dwarf galaxies, their specific spatial motion through the shocked environment seems to play a key role in determining whether they survive or quench. For example, galaxy 18 (Figure \ref{fig:g18}) has a stellar mass of $10^{8.5}$ M$_{\odot}$ ($M_{\rm gas}\sim10^{9.5}$ M$_{\odot}$ by $z=2$); yet, it exhibits a clear decline in sSFR after interacting with the shock, and its mass in bound gas stops increasing steadily, flattening after the shock interaction. The residual star formation after entering the shock-heated environment ($\sim0.6$ dex below the SFMS) appears to be sustained not by new gas accretion, but by the bound gas already accumulated, and it qualitatively appears that the system will likely quench after consuming this supply of gas, assuming it stays in the hot region. 

In contrast, galaxy 17, which is similar in mass, shows only a mild response in sSFR due to the shock. Examination of the postage stamps for this system over time indicates that from nearly the beginning of the simulation, this galaxy resides within a cold filament inside the sheet region. This filament appears to have shielded the galaxy from the heating caused by the shock, which does not penetrate the strongest of the filaments, for the majority of the simulation. It is worth noting that this is a reversal of the effect filaments have at low redshift, which we discuss further in Section \ref{discussion}.

The time of shock interaction also appears to play a role in whether systems enter our quenched sample; those that experience the shock late ($z\sim3$) tend to have enough cold gas to survive through the end of the simulation. Because the simulation ends at $z=2$, is it possible that many of these systems would suffer the same fate as those whose evolution we can track over longer periods due to their earlier shock interactions. Additionally, some systems which enter the shock at $z\lesssim3$ seem to be responding in ways similar to the quenched portion of the sample, but then enter within 3 \Rv{} of the protoclusters before they fully quench. In these cases we cannot attribute any further activity unambiguously to the sheet. 

\subsection{Effect on Lower Mass Dwarfs}

The lower mass end of the sample ($10^{5.5}<M_*<10^{7}$ M$_{\odot}$) exhibits a much more uniform response to the shock; nearly all show strong declines in mass accretion rate and sSFR as a result of the environmental change precipitated by the accretion shock, regardless of the specifics of their locations or paths through the region. In contrast with the more massive sample, a sizable fraction of these systems ultimately lose \textit{all} of their bound gas over the course of their interaction with the shock-heated sheet. These declines often occur despite there being a significant amount of gas present within $R_{\rm V}$ of the system, reflecting the inability of these systems to retain their gas, consistent as well with the increase in $T_{\rm amb} / T_{\rm V}$ to ratios well above unity. 

An illustrative example of this process is shown in Figure \ref{fig:173_shock}, which focuses on galaxy 173, selected because it was located far out in the IGM in a region devoid of other massive haloes. At $z\sim4.4$, the shock is rapidly approaching galaxy 173. As the shock hits and crosses over the system at $z\sim 4$, pockets of the region surrounding the galaxy rapidly rise over an order of magnitude in temperature. This can be seen in the large right panel, as well as the smaller zoom panels in Figure \ref{fig:173_shock}. By $z\sim 4$, the sSFR, gas mass accretion rate, and bound gas content of the galaxy are already declining, and by $z\sim2.5$, the galaxy is on the verge of quenching, with its bound gas fraction significantly depleted despite there being extant gas in its vicinity. The region highlighted in Figure \ref{fig:173_shock} is also a prime example of the ``shattering" of the sheet due to the shock, discussed in detail in \cite{Mandelker:2019} and \cite{Mandelker:2021}. 

These features are shown quantitatively for four systems in the sample spanning final stellar masses between $6<\log M_*<8$ M$_{\odot}$ (45, 61, 88, and 124) in Figure \ref{quenched_sample}.\footnote{Similar figures for the other galaxies in the sample are available in the online version or at http://ipm-dwarfs.github.io.} These galaxies all follow a similar pattern: as the shock progresses across the sheet and the ambient temperature of gas around the galaxies increases, we see a rise in $T_{\rm amb} / T_{\rm V}$ (panel 1), generally associated with a ``bump," or minor enhancement, in gas accretion onto their haloes superimposed on a broader long term reduction in accretion (panel 2). This bump is temporally correlated with the shock arrival, and likely results from gas compression as the shock hits the galaxy. These galaxies, prior to the shock, tend to have $T_{\rm amb} / T_{\rm V} \simeq 1$, indicating that for their mass and the pre-shock ambient temperature of the sheet region, they are marginally able to retain their bound gas. 

As we track the gas content of these galaxies past this point (panel 3), the amount of bound gas begins to flatten or decrease, in some cases dropping to zero. Finally, these processes are reflected in the specific star formation rate of these systems (panel 4), which after the arrival of the shock and continued heating of the ambient gas, tends to fall, either rapidly or more steadily, to the quenching threshold 1 dex below the main sequence. We see on occasion that the arrival of the shock and accompanying enhancement in gas accretion triggers an isolated enhancement in sSFR before the decline. This gas compaction effect has been noted before for ram pressure stripped galaxies in clusters \citep[e.g.][]{Ricarte:2020}, albeit at higher mass than we focus on here. 

Of the galaxies with final stellar mass less than $10^{7}$ M$_{\odot}$, all but three quench as a result of the shocked sheet. Two of the three that do not strictly meet our quenching definition (92 and 115) both drop past the quenching threshold and oscillate in sSFR around it; as noted in Table 1, roughly half of the final 20 snapshots of the simulation for each find them below the threshold. While we don't strictly define these two systems as quenched, they qualitatively match the behavior of the other low mass systems in the sample. 

Galaxy 140 mirrors the behavior of galaxy 18, dropping to a new reduced sSFR but not quenching over the course of the simulation. Based on the postage stamps, it appears to be the lone example in the lower mass subsample to be shielded by a cold filament for much of the simulation.

As a point of contrast, Figure \ref{outofsheet} shows a galaxy which evolves in the simulation \textit{outside} the shocked sheet, only falling into one of the massive clusters in the last several snapshots of the simulation. Unlike the in-sheet systems discussed, the $T_{\rm amb} / T_{\rm V}$ remains extremely steady at unity until cluster infall, with steady, monotonic growth in bound gas and stellar mass.

\section{Discussion}\label{discussion}
\subsection{Role of the IPM Environment in Quenching}
In the \hypa{} simulation, which covers redshifts from $z\sim7$ to $z\sim2$, the ultimate fate of many of the dwarf galaxies which quench as centrals is to become satellites of a proto-cluster or another more massive halo. Because these systems quench before becoming satellites, they are examples of pre-processing in effect \citep[e.g.,][]{Balogh:2000,Wetzel:2013}. It is important to distinguish this form of pre-processing from that which is inferred at low redshift; we find that at high-$z$, dwarf galaxies in IPM regions that are more isolated (particularly from filaments) tend to preferentially quench due to the large scale accretion shock (e.g., galaxy 173 or 61), while those which may be in the IPM but are within filaments tend to survive the shock and continue forming stars (e.g., galaxy 17 and 140). This is the opposite behavior than is seen at low redshift, where filaments are hot, and thus tend to quench dwarfs before their infall into clusters \citep{Dekel:2006,Birnboim:2016}. Thus, the high$-z$ IPM environment, when undergoing such shocks, acts in the role held by galaxy groups and filaments at low$-z$ in driving quenching, while high$-z$ filaments (which tend to be cold) have the opposite effect from their low$-z$ counterparts. Interestingly, evidence has also been found observationally of the pre-processing of more massive systems before cluster infall at $z\sim1$ ($M_*>10^{10}$ M$_{\odot}$) \citep{Werner:2021}. It thus seems likely that pre-processing can have effects on infalling galaxy populations across a range of mass scales.

While the fate of most of the systems identified in this work is cluster infall, it is reasonable to infer that similar sheet-shocking events at lower redshift have the capacity to generate field dwarfs that are quenched. To date, studies at low redshift have found no evidence of such systems statistically \citep[e.g.,][]{Geha:2012}; however, the volumes probed by such studies in the low mass regime ($M_{*}<10^{8}$ M$_{\odot}$) are thus far relatively small due to these objects' low surface brightnesses. A handful of individual dwarf galaxies in relative isolation have been found in states of quiescence \citep[e.g.,][]{Makarov:2012,Karachentsev:2015,Polzin:2021}, but upcoming and ongoing deep, low surface brightness surveys such as the Dragonfly Wide Field and Ultra Wide surveys \citep{Danieli:2020} and DESI Legacy Imaging Survey \citep{Dey:2019}, and the Legacy Survey of Space and Time \citep{lsst} over its longest baseline are poised to uncover new populations of low surface brightness dwarf galaxies in the $10^5$ to $10^8$ M$_{\odot}$ mass range, some of which may be quenched centrals in relative isolation having experienced a shocked sheet environment during their evolution.

\begin{figure}
    \centering
    \includegraphics[width=\linewidth]{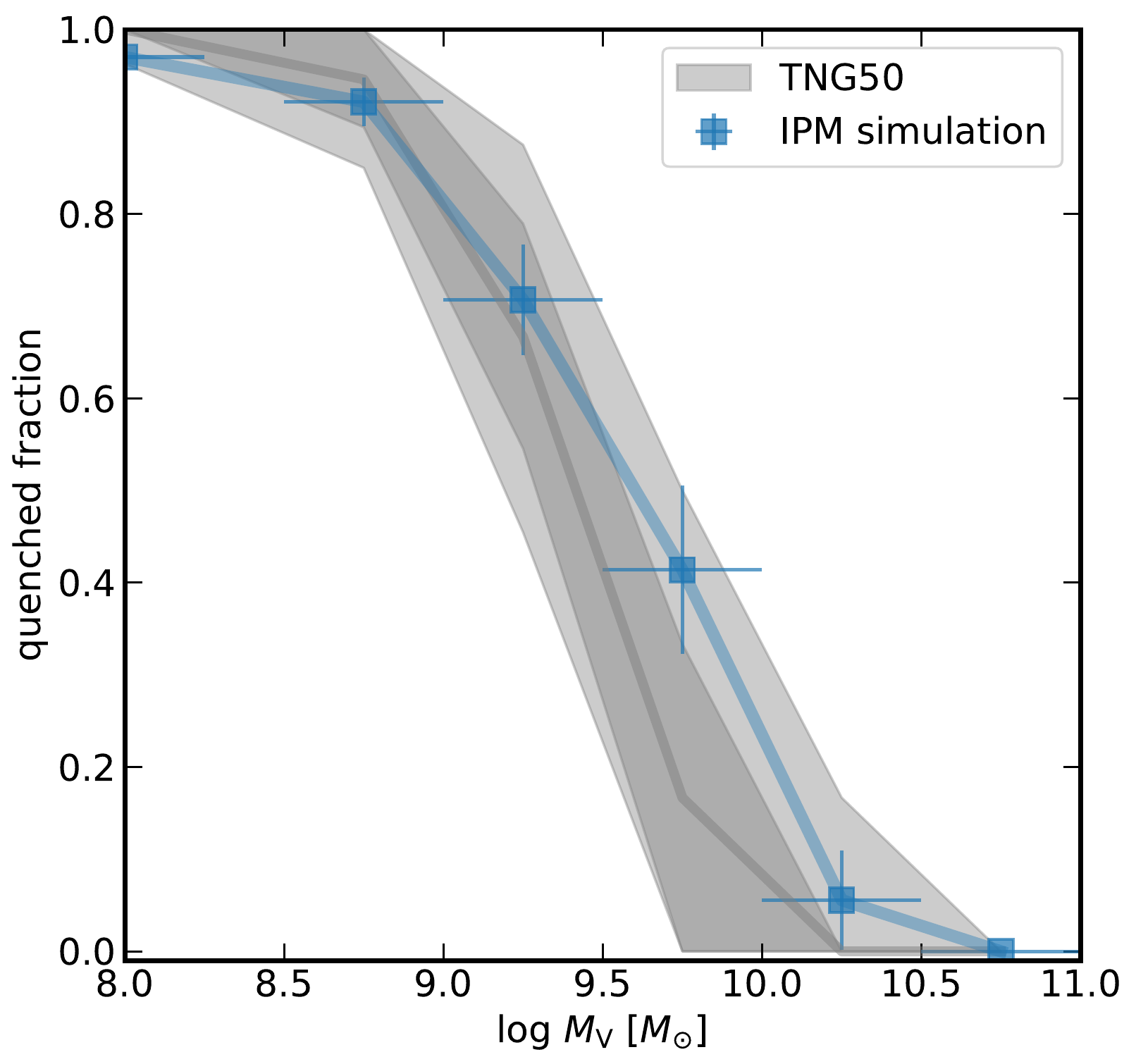}
    \caption{Comparison between quenched fractions in the \hypa{} sheet at $z=3$ and those in TNG50 subvolumes selected as analogues to the \hypa{} region. The dark grey line shows the median quenched fraction for each mass bin across the TNG50 sample, while the gray filled region shows the 16-84 (dark grey) and 5-95 percentile (light gray) range. The trends seen are in reasonable agreement with the quenched fraction recovered from the \hypa{} simulation region at this redshift. There is evidence of higher quenched fractions in \hypa{}, particularly in the $9<\log M_{\rm V}<10$ bins. Such a trend is expected given that \hypa{} is a biased region which we have shown to preferentially lead to quenching.}
    \label{fig:TNG50_qf_full}
\end{figure}

\subsection{Comparison with TNG50}
A key question in the interpretation of these results is the degree to which accretion shocks due to large scale structure collapse are ubiquitous in the universe, and how often these are associated with thin, well defined two dimensional sheets which allow for these shocks to cover a large area. While this cannot be extrapolated from \hypa{}, it is worth noting that large scale hydrodynamic simulations predict that up to half of the Universe's baryon content could be in the form of shock-heated gas in the intergalactic medium \citep{Cen:2006}. 

The TNG50 simulation \citep{Nelson:2019} was carried out with the same physics as both TNG100 and \hypa{}, and represents an avenue for exploring this question, particularly given the galaxy masses targeted in this work. In forthcoming work, we examine the full TNG100 dataset for sheet regions and large scale accretion shocks across cosmic time and investigate their prevalence, finding that shock-heated sheets and cold co-planar filaments, with similar properties to those seen in \hypa{}, are quite common around massive haloes with $M_{\rm V}\sim 10^{12}$ M$_{\odot}$ at $z>2$ (Flynn et al. \textit{in prep}). While a detailed convergence analysis is beyond the scope of this paper, we note that when comparing the global quenched fractions away from large haloes between TNG100, TNG50, and the four zoom factors of \hypa{}, we find that generally, ZF3 and ZF4 are in agreement and converged, and TNG50 is roughly in agreement with ZF2, and represents a considerable improvement over TNG100 and ZF1.

In this section, we aim to make as close to an ``apples-to-apples'' comparison to TNG50 as is feasible to determine whether the quenched fractions seen in the shock heated sheet in \hypa{} are consistent with similar regions throughout the TNG50 volume. Because our results are sensitive to region to region variations, e.g., cosmic variance, we do not treat the TNG50 volume as a single region; instead, we subdivide the TNG50 volume into subvolumes which are approximately the size of the volume tracked in \hypa{} (2000 by 2000 by 4000 kpc at $z=3$). This produces $\sim$6000 subvolumes in TNG50 (depending on the orientation chosen --- due to the elongation of the boxes and the buffers chosen at the edges of the simulation, the different orientations of the boxes produce similar but not identical numbers of boxes).

We wish to compare \hypa{} to volumes that at least qualitatively meet some similarity requirements with \hypa{} --- we choose to select subvolumes which contain at least a small population of low mass dwarf galaxies (7.5$<\log M_{\rm V}<$8.5) that have nonzero stellar mass. We set the minimum population size to 15, but any value between 5 and 20 produced nearly the same subvolumes. Roughly 700-800 of the subvolumes in TNG50 met this requirement, again dependent on orientation. We then computed the quenched fractions, ignoring systems within 3 $R_{\rm V}$ of a massive halo ($>10^{11}$ M$_{\odot}$) in each subvolume. Next, we repeated the experiment by rotating the long axis of our boxes to two other orthogonal orientations. This ultimately produced a sample of 2179 subvolumes which all sample the TNG50 simulation on a small spatial scale, and contain a given system $\sim3$ times. We consider this our ``regional bootstrapping" sample, and compute the mean quenched fractions as in \hypa{}, estimating the box-to-box variance from percentiles across the full sample of 2179. We note that using any single orientation instead (e.g., $\sim700-800$ boxes) produced the same results. Occasionally, an individual mass bin in an individual subvolume contained no galaxies meeting our criteria; these were not included. 

\begin{figure}
    \centering
    \includegraphics[width=0.95\linewidth]{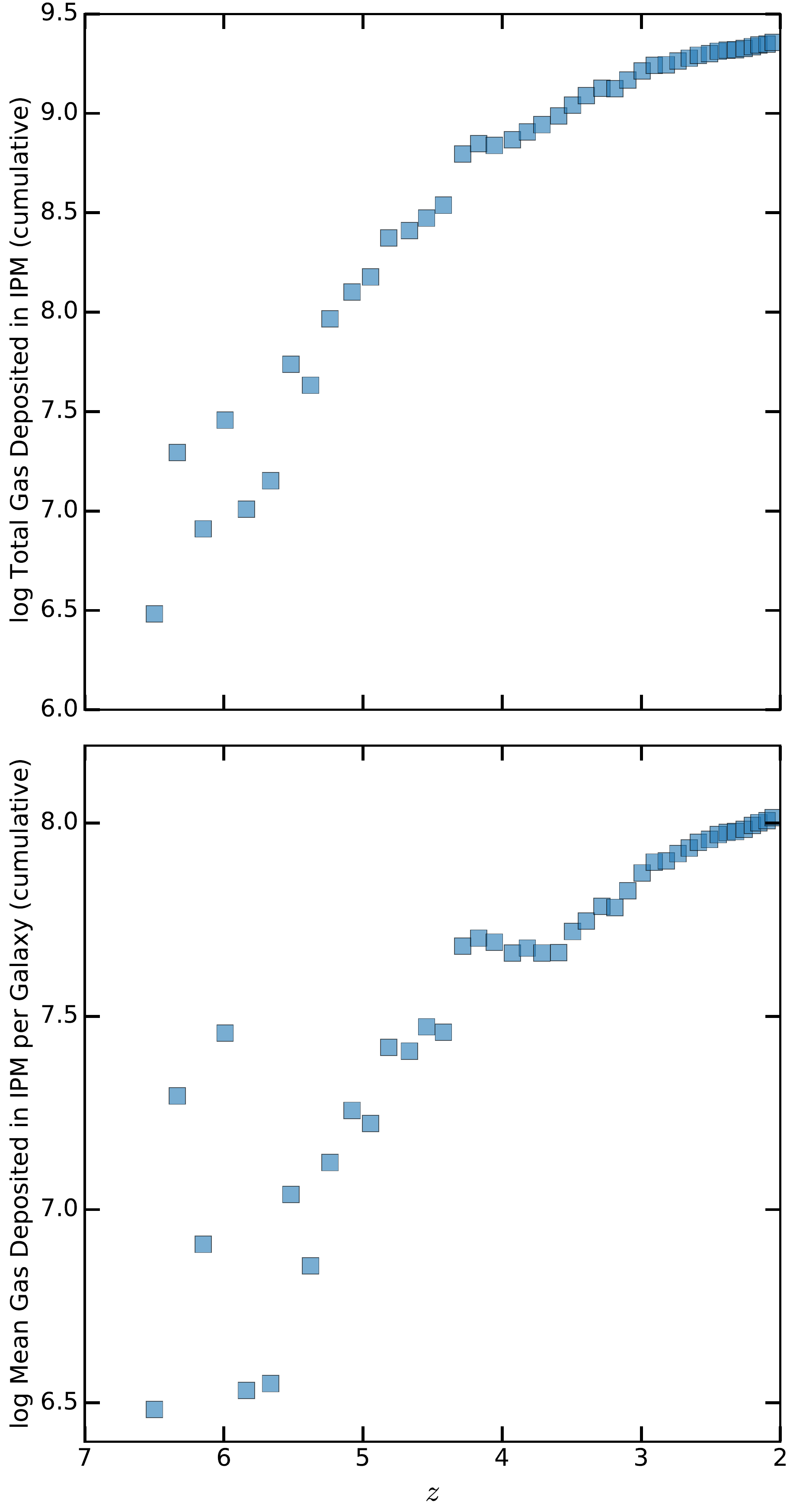}
    \caption{Cumulative bound gas deposited into the IPM from galaxies in our sample, both total (top) and average per system (bottom). We find that the amount of gas liberated from these systems is a negligible contribution to the HI clouds noted by \citet{Mandelker:2021}; in particular, while that work found that the amount of HI in the IPM was $\sim 5\times10^{10}$ M$_{\odot}$ at $z=4$ (more than an order of magnitude than was deposited by that point here), the gas deposited by stripped dwarf systems is likely predominantly hot, having been liberated primarily from the CGM.}
    \label{deposited}
\end{figure}

The result of this exercise is shown in Figure \ref{fig:TNG50_qf_full}. We find that there is good qualitative agreement between the TNG50 subvolumes and the \hypa{} simulation, with our quenched fractions lying within the 5-95 percentile range of calculated values. There is some evidence of increased quenching in the $9.0<\log M_{\rm V}<10.5$ mass range in \hypa{} over the TNG50 mean; this is consistent with an interpretation that the sheet region studied is over-quenched compared to an ``average" region in the universe, but is still within the spread of similar-sized regions found in the larger simulation. It is also consistent with the virial mass range of the quenched sample discussed in this work. The extension of this interpretation is that the subvolumes in TNG50 driving the upper boundary of the quenched fraction spread are likely to contain, e.g., shock-heated sheets or other similar structures. Further investigation is necessary to explore this more systematically in cosmological simulations. We note that resolution effects make the observed difference stronger --- ZF2, which is roughly matched in resolution to TNG50, is overquenched by $\sim5$\% with respect to the ZF3 curve shown (particularly in the $9<\log M_{\rm V}<10.5$ M$_{\odot}$ bins). Put another way, if TNG50 were run at \hypa{} resolution, we would expect the measured quenched fractions to be suppressed somewhat with respect to those shown in Figure \ref{fig:TNG50_qf_full}.

\subsection{Stripped Gas Deposited into the IPM}
The general evolution of low mass galaxies in the IPM tends to involve the loss of a sizable fraction of their bound gas mass, which is deposited back into the IGM in the sheet. While this is not expected to be an outsized affect on the multiphase structure of the IPM, it is worth considering whether the amount of gas liberated from all systems in our sample that interact with the shock-heated IPM is enough to have a bearing on the interpretation of the results of \cite{Mandelker:2021}, in which $\sim5\times10^{10}$ M$_{\odot}$ of HI gas, and roughly an order of magnitude more total Hydrogen, was found within the IPM region at $z=4$. 

For the galaxies in our sheet sample, we compute the amount of gas liberated from each system between its point of maximal bound gas and the end of the simulation, accounting for any star formation between those points. We then construct the cumulative and average gas depositions into the IPM from these systems (Figure \ref{deposited}). We find that on average, each galaxy in our sample contributes $\sim 10^8$ M$_{\odot}$ of gas to the IPM after interacting with the shock, with a total of $\sim 2\times10^9$ M$_{\odot}$ of gas being contributed by all systems in our sample. Expanding to the larger number of sheet-located dwarfs (those which didn't make our sample due to stellar mass cuts), this roughly estimates 6$-$7 $\times 10^{9}$ M$_{\odot}$ of deposited gas.

At $z=4$, we find that only $\sim10^{9}$ M$_{\odot}$ of gas has been stripped from our sample of dwarf galaxies in the IPM. Furthermore, we find qualitatively that gas is primarily liberated from the hot/CGM phase, and thus does not \textit{directly} contribute to the HI masses quoted in \cite{Mandelker:2021}, though whether this stripped gas precipitates any further thermal instabilities is an open question. As we cannot effectively track direct stripping of ISM gas into the IPM, versus gas that passes through the CGM, these results constitute primarily an upper limit. It thus seems that stripping of bound gas from galaxies in the IPM accounts for \textit{at most} $\sim10$\% of the total HI content. Assuming each galaxy contributes a generous $10^8$ M$_{\odot}$ of gas, $\sim$5000 systems would be needed to contribute the total amount of gas measured in \cite{Mandelker:2021}, which is far greater than the number of relevant-mass systems present in the shocked region ($N<100$). It is thus most likely this HI gas does not originate from (stripped) galaxies, but rather formed directly out of the baryonic material in the
sheets due to thermal instabilities.

\section{Summary and Conclusions}\label{summary}

Using the high resolution zoom-in simulation \hypa{}, which provides the highest resolution of a large patch of the IGM to date, we have identified a sample of dwarf galaxies (at $z=3$) which are not satellites and which live in a cosmic (or Zel'dovich) sheet in between two massive halos, aka the intrapancake medium (IPM). Between $z=5$ and $z=3$, two cosmic sheets in the simulation collide, triggering an accretion shock which propagates across the simulation volume and shock heats the IGM by over an order of magnitude in temperature. Many of the dwarf galaxies in the sheet region had virial temperatures $T_{\rm v}$ --- which traces how well a system can hold on to its gas --- that were only marginally greater than the ambient temperature of the IGM before the shock. The arrival of the shock to the vicinity of many of these systems had a multifold effect, including a brief enhancement in gas accretion (and occasionally in sSFR as a result), followed by a rapid increase in ambient temperature (and corresponding rise in $T_{\rm amb}$ / $T_{\rm V}$) which generally prevented further gas accretion and suppressed star formation. Over time, many of these systems then lost some, or all, of their bound gas to environmental effects such as ram pressure stripping and starvation. This often led ultimately to the galaxy's quenching. 

The strength of the aforementioned process varies from galaxy to galaxy, driven primarily by initial mass and individual environmental differences as each moves through the sheet. We find several examples of dwarf galaxies with $M_{*}>10^{7}$ M$_{\odot}$ that quench due to this process, and galaxies as massive as $M_{*}>10^{8.5}$ M$_{\odot}$ that show a definite decrease in sSFR due to the environment of the shock-heated sheet. For galaxies with $M_{*}<10^7$ M$_{\odot}$, quenching due to the hot sheet is almost guaranteed. 

This mechanism, if such accretion shocks on cosmological scales are common, provides a pathway for dwarf galaxies to quench in ``isolation'' --- they need not interact with a more massive halo. This quenching is not driven by feedback or other internal processes (which are thought generally to be unable to permanently quench galaxies in this mass range); rather, the large scale environment of the sheet mimics the environments traditionally considered drivers of quenching, such as clusters, galaxy groups, and the halos of massive galaxies.

How common such systems might be, whether as pre-processed satellites or as quenched dwarfs in the field, depends on the frequency of sheet regions, the number density of susceptible dwarf galaxies that live in sheet regions, and the frequency of large scale structure collapse events which propagate shocks. None of these parameters are currently well constrained. Future work will aim to generate a census of sheet regions in cosmological simulations such as TNG50 and TNG100, and to track their evolution over cosmic time to evaluate the frequency of large scale shocks. 

\section{ACKNOWLEDGEMENTS}
The authors thank Ruediger Pakmor for helpful discussions and for their considerable aid in getting the simulation I/O scripts installed on a problematic compute node. This work was partly funded by the Klauss Tschira Foundation through the HITS Yale Program in Astrophysics (HYPA). I.P. is supported by the National Science Foundation Graduate Research Fellowship Program under grant No. DGE1752134. NM acknowledges support from the Gordon and Betty Moore Foundation through Grant GBMF7392 and from the National Science Foundation under Grant No. NSF PHY1748958. FvdB is supported by the National Aeronautics and Space Administration through  Grant No. 19-ATP19-0059 issued as part of the Astrophysics Theory Program. FvdV is supported by a Royal Society University Research Fellowship.

\section{Data Availability}
The data underlying this article will be shared on reasonable request to the corresponding author.

\bibliography{example}{}
\bibliographystyle{mnras}

\bsp	
\label{lastpage}
\end{document}